\documentclass[acmlarge,screen]{acmart}

\setcopyright{acmcopyright}
\copyrightyear{2020}
\acmYear{2020}

\acmMonth{5}

\usepackage{soul}
\usepackage{url}
\usepackage{amsmath}
\usepackage{amsthm}
\usepackage{xcolor}
\usepackage{algorithm}
\usepackage[noend]{algpseudocode}

\newcommand{\apt}[1]{\textcolor{black}{#1}}
\newcommand{\final}[1]{\textcolor{black}{#1}}
\newcommand{\ncz}[1]{\textcolor{black}{#1}}

\newcommand{\sysname}{WiFiTrace }

\newcommand{\sysnames}{WiFiTrace}

\title{\sysnames: Network-based  Contact Tracing for Infectious Diseases Using Passive WiFi Sensing}

\author{Amee Trivedi }
\affiliation{University of Massachusetts Amherst }
\email{amee@cs.umass.edu }
\author{Camellia Zakaria }
\affiliation{University of Massachusetts Amherst }
\author{Rajesh Balan }
\affiliation{Singapore Management University }
\author{Ann Becker }
\affiliation{University of Massachusetts Amherst }
\author{George Corey }
\affiliation{University of Massachusetts Amherst }
\author{Prashant Shenoy }
\affiliation{University of Massachusetts Amherst }

\renewcommand{\shortauthors}{Trivedi et al.}

\begin{document}

\ccsdesc[500]{Human-centered computing~Ubiquitous and mobile computing systems and tools}

\keywords{Digital Contact Tracing; Passive sensing; WiFi; Access Point; Syslogs;}

\setcopyright{acmlicensed}
\acmJournal{IMWUT}
\acmYear{2021} \acmVolume{5} \acmNumber{1} \acmArticle{37} \acmMonth{3} \acmPrice{15.00}\acmDOI{10.1145/3448084}

\begin{abstract}
Contact tracing is a well-established and effective approach for the containment of the spread of infectious diseases. While Bluetooth-based contact tracing method using phones has become popular recently, these approaches suffer from the need for a critical mass adoption to be effective. In this paper, we present WiFiTrace 
, a network-centric approach for contact tracing that relies on passive WiFi sensing with no client-side involvement. Our approach exploits WiFi network logs gathered by enterprise networks for performance and security monitoring, and utilizes them for reconstructing device trajectories for contact tracing. Our approach is specifically designed to enhance the efficacy of traditional methods, rather than to supplant them with new technology. We designed an efficient graph algorithm to scale our approach to large networks with tens of thousands of users. The graph-based approach outperforms an indexed PostgresSQL in memory by at least 4.5X without any index update overheads or blocking. We have implemented a full prototype of our system and deployed it on two large university campuses. We validated our approach and demonstrate its efficacy using case studies and detailed experiments using real-world WiFi datasets.

\end{abstract}

\maketitle

\section{Introduction}
\label{sec:introduction}

Seasonal Influenza-like Illnesses (ILI) have become prevalent in our society, with nearly 45 million illnesses in 2018 alone \cite{flu2008estimated}. 
The emergence of coronaviruses such as SARS in 2002~\cite{centers2016sars} and COVID-19 in 2020, which are more infectious and lethal than the flu, have added to the urgency for developing better tools for managing containment and mitigation strategies. Many computational and data-driven tools have been designed to address these needs in recent years, including epidemiological models~\cite{salathe2010high}, flu prediction tools~\cite{reich2019collaborative}, visualization tools, and even embedded sensing devices \cite{FluSense}.

In this context, contact tracing, combined with testing, has emerged as a particularly effective containment and mitigation method~\cite{salathe2020covid,klinkenberg2006effectiveness,hellewell2020feasibility}. Contact tracing involves identifying who may have come in contact with a person who has contracted an infectious disease, with a recursive tracing of their contacts~\cite{eames2003contact}. Countries like Singapore, South Korea, and Germany have demonstrated the efficacy of aggressive contact tracing during the COVID-19 pandemic~\cite{taaa039,Korea}. While manual methods for contact tracing are well established in the medical community, such methods are laborious and do not scale well. As a result, computational methods based on phone sensing have received significant attention recently. Technology companies such as Apple and Google recently announced phone-based contact tracing support in Android and iOS~\cite{Apple_Google}, and numerous mobile phone apps for contact tracing have been deployed by the public health agencies and health researchers~\cite{ferretti2020quantifying}. These methods assume that smartphones are ubiquitous and carried by their owners everywhere they go, allowing phone sensing methods to detect proximity to other users. These client-side approaches use Bluetooth for proximity sensing, sometimes combined with GPS and other localization techniques present on the phone for location sensing~\cite{bay2020bluetrace}.

In this paper, we present \emph{\sysnames}, an alternative network-centric approach for phone-based contact tracing. In contrast to client-side approaches that depend on Bluetooth and mobile apps, a network-centric approach does not require on-device data collection or apps to be downloaded on the phone. Instead, users use their phone or mobile device normally, and the approach uses the {\em network's view of the user} to infer their location and proximity to others. Our approach is based on WiFi sensing~\cite{JaisinghaniBNML18,Zhou2015} and leverages data such as system logs (``syslogs'') generated by the enterprise WiFi networks for contact tracing. Although our approach does not require WiFi location~\cite{mok2007location}, such techniques, where available, can further enhance the efficacy of our approach. Our network-centric approach to contact tracing offers a different set of tradeoffs and privacy considerations than Bluetooth-based client-centric methods; one goal of our work is to \ncz{carefully} analyze these tradeoffs. Note: \sysname was developed in collaboration with key healthcare professionals, including the director of a campus hospital and a nurse who have been running contact tracing operations for various outbreaks (such as meningitis) for many years.

The following scenario presents an illustrative use case of how our approach works. Consider a student who visits the university health clinic and is diagnosed with an infectious disease. The university health clinic officials decide to perform contact tracing and seek the student's consent for network-based contact tracing. Because the infected student could have transmitted the disease to others over the past several days, it is important to determine the set of other students who spent sufficient time in the affected student's proximity at any place on campus in the last few days. To do this, the health officials input the WiFi MAC address of the infected student's phone into \sysnames. It then analyses WiFi logs generated by the network, specifically association and dissociation log messages for this device, at various campus access points to reconstruct the locations (building, room numbers) visited by the user. It analyzes all other users associated with those access points to determine users in the proximity of the infected student and duration of co-location. These locations and proximity reports are then used by health officials to assist with contact tracing. Additional reports for each impacted user can be recursively generated. 

Our paper makes the following contributions:
\begin{itemize}
    \item We present \sysnames, a network-side contact tracing method that involves passive WiFi sensing and no client-side involvement. We discuss why such an approach may be preferred in some environments, such as academic or corporate campuses, over client-side methods.
    \item We present a graph-based model and graph algorithms to efficiently perform contact tracing on passive WiFi data comprising tens of thousands of users. We show that our graph-based approach can scale to settings with tens of thousands of users. We show that the graph-based approach achieves similar performance to an indexed PostgresSQL database but with 4.5X less memory usage and without any index update overheads or blocking. 
    \item We implemented \sysname and deployed it at two large university campuses on two different continents.
    \item We validated and experimentally evaluated our approach using anonymized data from two large university networks. Our results show the efficacy of \sysname at contact tracing for three simulated diseases. These results also highlight the need to judiciously choose WiFi session parameters to reduce both false positives and false negatives. Overall, our validation results show that \sysname can accurately record all user sessions that last for 3 minutes or more. 
    
\end{itemize}

\section{Background}
\label{sec:background}

In this section, we provide background on contact tracing and present motivation for our network-centric approach.

\subsubsection*{\textbf{Contact Tracing Procedure:} } Contact tracing is a well-established method used by health professionals to track down the source of an infection and take pro-active measures to contain its spread~\cite{eames2003contact}. The traditional method is based on questionnaires, whereupon diagnosis, the user is asked to list places visited, and other people they have contacted. This information is used to contact the set of potentially infected individuals, and the process is continued recursively until all possible infections are eliminated~\cite{eames2003contact}. The contact tracing goals are two-fold: identify the potential infection source for the diagnosed individual and determine others who may have gotten infected due to proximity or contact.

Unfortunately, many illnesses have a  2 to 14 day incubation period between infection time and the onset of the illness. Thus, infected users might need to recall the places they had been to and the people they interacted with two weeks ago. The reliance on human memory to trace the previous location and interaction footprints is a laborious manual process and erroneous. 


\subsubsection*{\apt{\textbf{Client-centric Digital Contact Tracing:}}} \apt{With the ubiquity of smartphones, digital contact tracing leveraging Bluetooth and GPS~\cite{bay2020bluetrace,Apple_Google} has become popular. In this approach, a unique (often anonymized) identifier is transmitted, using Bluetooth, from every phone. Each phone also listens for such identifiers from other phones to determine users/phones in their proximity. This information is further used for contact tracing.
Apple and Google have implemented this basic approach as part of their contact tracing API~\cite{Apple_Google}, which has been utilized by many standalone apps~\cite{PanEurope,COVIDSAFE, safe_paths, bay2020bluetrace, trace-together}.}

\apt{These solutions are client-centric and require high user adoption for successful contact tracing coverage. In particular, each user must download the standalone app, allow the necessary permissions to sense, and allow the app to run continuously in the background. The need for high user adoption proved to be a key hurdle for Singapore's TraceTogether app~\cite{trace-together}, which achieved only 1.1 million downloads despite needing a critical mass of 4 million active users (around two-thirds of the population) to be effective~\cite{WSJ-TT}. These challenges have led health experts to argue that while technology-based contact tracing solutions are useful, they should be seen as complementary yet effective to traditional \ncz{contact tracing methods}~\cite{ferretti2020quantifying}.
}

\subsection{Motivation of \sysname} 

\subsubsection*{\apt
{\textbf{Aid Health Experts:}}} \apt{ \sysname was designed to support health experts performing contact tracing in two key ways. First, \sysname aids in acquiring detailed location histories of individuals for whom contact tracing is necessary. This significantly reduces the burden and errors caused by high dependence on human recollection to determine potentially infected contacts. Second, \sysname processes the overwhelming amount of location information to prioritize co-locators at high risk of infection. This significantly reduces the number of contacts that \ncz{need follow-up} by the health experts. In addition, unlike standalone contact tracing apps that require end-users to self-monitor and self-report their proximity to infected users, \sysname is designed to integrate with a health expert's contact tracing workflow.}


\begin{figure*}[h]
\includegraphics[scale=0.7]{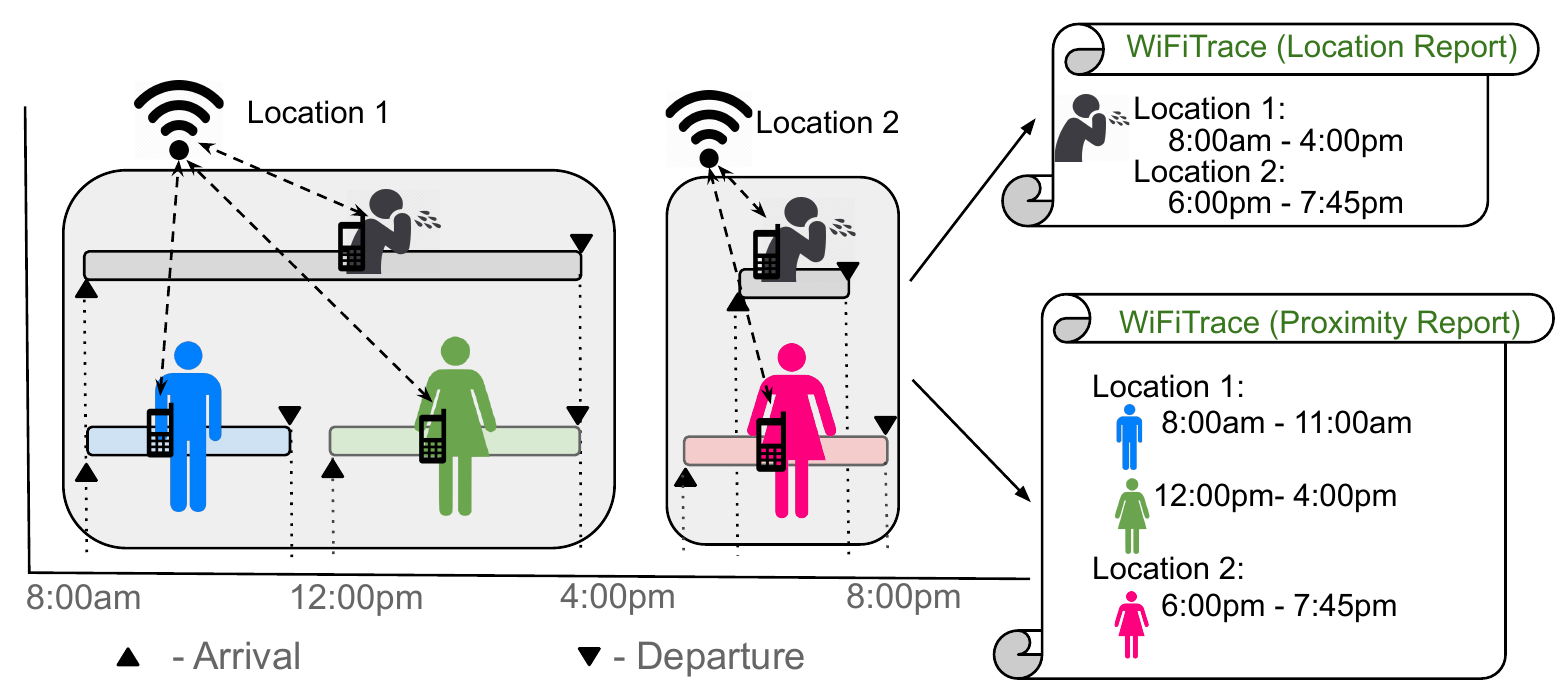}
    \caption{A network-centric approach to contact tracing using WiFi sensing
    \label{fig:wifi_contact_tracing}}
\end{figure*}

\subsubsection*{\apt{\textbf{Network-centric Digital Contact Tracing:}}} \apt{For \sysname to be useful to health experts, it must overcome the key adoption challenge faced by client-centric approaches. To achieve a high adoption rate, \sysname uses a network-centric sensing approach using passive WiFi sensing, as illustrated in Figure \ref{fig:wifi_contact_tracing}.}

These days, WiFi access is almost ubiquitous in key environments such as offices, university campuses, and shopping areas. WiFi sensing has thus emerged as a popular approach for addressing a range of analytic tasks~\cite{JaisinghaniBNML18, Zhou2015}. WiFi sensing can be client-based (i.e., done on the mobile device) or network-based (i.e., done from the network's perspective). An example of client-side WiFi sensing is performing triangulation via RSSI or time-of-flight measurements to multiple WiFi access points and localize a device's position~\cite{mok2007location}. 

On the other hand, network-centric WiFi sensing uses the network's view of where individual WiFi devices are located to perform analytics. This approach has been used for monitoring the mobility of WiFi devices by analyzing the sequence of the access points that see the same device over a period of time~\cite{JaisinghaniBNML18}. 

WiFi sensing (both client and network-centric) has been used to characterize and model user mobility~\cite{kotz2005analysis,kim2006extracting} with more recent work leveraging WiFi sensing to track health~\cite{Khan2017}, stress~\cite{Zakaria19}, and perform retail analytics~\cite{Zeng15}. 

We build on prior work and leverage a network-centric WiFi sensing approach to assist with contact tracing. Our key insight is that the {\em mobility of a user's phone is visible to the network} through the sequence of WiFi access point associations performed by the device as the user moves. This allows the network to determine the locations visited by the users' device and other co-located devices present at those locations by being associated with those APs. Thus, the approach passively senses devices as they move through the WiFi network. The key advantages of this approach over a client-centric approach are:

\begin{enumerate}
    \item \apt{\textbf{No mass user adoption:} Our approach does not require users to opt-in or download an app as the wireless network will automatically ``see'' all connected devices (associated) to it at all times. This approach makes \sysname much easier to deploy at scale.}
    
    \item \apt{\textbf{No active data collection on user devices:} Data is collected directly by the WiFi network without requiring any direct interaction with a user's device (unlike a client-centric approach). Specifically, our method relies on collecting ``syslog'' network events, SNMP reports, or RTLS events that are already commonly used by many enterprise networks for performance and security monitoring.}
    
    \item \apt{\textbf{Single sensing modality:} A client-centric method that uses Bluetooth for proximity sensing must use a second sensing modality such as GPS for sensing device location. In contrast, WiFi sensing is single modality sensing determining both location (based on AP locations) and proximity (based on AP associations). It is important to note that GPS does not work well indoors while WiFi does. 
    }
    
\end{enumerate}

\subsubsection*{\apt{\textbf{Design Challenges:}}} \apt{Our network-centric approach is not without several challenges. Firstly, WiFi sensing only provides coarse-grain proximity measurements (e.g., users co-located within the range of an access point). This is unlike fine-grain Bluetooth sensing, which promises accuracy values of up to a few feet \ncz{from} the user. Thus, using coarse-grained proximity sensing could increase the number of false positives. To overcome this, \sysname capitalizes on the co-location duration as an indicator of the risk of infection to investigate individuals at higher risk \footnote{Traditional contact tracing can then process this candidate set to eliminate those who are not at risk.} as demonstrated in the evaluation Section \ref{sec:evaluations}. }


\apt{Secondly, WiFi-based contact tracing is limited to areas with WiFi coverage, such as indoor spaces and bounded outdoor spaces. In contrast, Bluetooth sensing does not need a network to listen to other devices and works ``anywhere''. We acknowledge this limitation of our network-centric approach but note that this approach is still highly effective in environments that do provide WiFi. In particular, these effective environments include most university campuses and corporate offices -- i.e., places where individuals spend a significant portion of their day. Table~\ref{tab:comparison} summarizes the comparison between a Bluetooth based client-centric approach and a WiFi-based network-centric approach.} 

\begin{table}
\caption{Comparison of Bluetooth as a client-centric vs. WiFi as a network-centric approach to contact tracing.}
\begin{tabular}{lll}
\hline
             & Bluetooth & Passive WiFi \\ \hline
 Architecture & Client-based & Network-based\\
 Location sensing& GPS & AP-level or WiFi locationing\\
Proximity sensing & Bluetooth & AP-level co-locators\\
 Distance sensing & Fine-grain & Coarse-grain \\
 Proximity duration & Fine-grain & Fine-grain \\
 Data collection & On-device & In-network \\
 Target environment & Indoors or outdoors & Indoors, limited outdoors\\
 Key technical hurdle & Mass adoption needed & Does not work outdoors\\
 Privacy Issues & Yes (see \cite{cho2020contact}) & Yes (See Sec \ref{sec:privacy}) \\ \hline 
\end{tabular}
\label{tab:comparison}
\end{table}

\subsection{\apt{Privacy and Ethical Considerations}}
\label{sec:privacy}
\apt{Client-centric contact tracing apps raise important privacy concerns, especially for everyday users whose whereabouts would essentially be documented~\cite{cho2020contact}. Network-centric WiFi-based sensing bears the same concern, especially since users are automatically and passively tracked when connected to the WiFi network. However, unlike most Bluetooth apps that are focused on assisting the population at large, our tool aims to assist case investigators, as our primary end-users, in performing the formal contact tracing procedure. To reduce \final{privacy} risk, several safeguards can be put into practice.}
\begin{enumerate}
    \item \apt{\textbf{No direct access to the user or device data:} Unlike client-centric contact tracing apps, users do not need to provide their information or change device permissions to provide apps access to device data. In our work, all WiFi network data such as MAC ID and username that can identify a user was hashed (e.g., using the SHA-2 hash) to maintain anonymity.} 
    
    
    \item \apt{\textbf{Leveraging existing operational security standards:} WiFi network data is used by many IT departments for network maintenance and security surveillance. For example, our campus uses the same WiFi data used by \sysname to track down compromised devices that may be responsible for internal DDOS attacks and identify student hackers who, most notably, might be attempting to change course grades. Additionally, in many regions, audit and compliance laws necessitate gathering network logs for analysis and audits. These routine evaluations have operational security standards in place to protect user privacy. Using this WiFi data for contact tracing requires compliance with the same high operational security standards.}
    
    \item \apt{\textbf{Emergency disclosure request:} A larger outbreak of a disease such as COVID-19 will require de-anonymization of location histories of high-risk individuals. This information will strictly be shared with case investigators performing the contact tracing. A typical procedure for \sysname will be to query only hashed identities and MAC addresses. Note that the hashed data is stored separately from the tool and only accessible to a small, trusted group. When an individual is identified as at-risk, only an authorized case investigator can obtain a de-anonymized copy of the information.}
    
    \item \apt{\textbf{Obtain user consent:} Data protection acts enacted in many countries require organizations to acquire user consent before starting any data collection operations. Similarly, before \sysname can be used in production, users must provide informed consent to contact tracing upon connecting to WiFi in an enterprise network. Similarly, case investigators must be authorized to retrieve de-anonymized information when necessary, and all data sharing must follow the approved guidelines. For example, a contact tracing team could decide to directly contact an individual at high risk of contracting COVID-19 or publish at-risk locations like a public alert to \final{appeal to potentially infected individuals to contact health authorities}. In the latter case, the proximity data report is used for further contact tracing when at-risk individuals contact health officials. We currently use the latter approach at our USA campus.}

\end{enumerate}

\subsubsection*{Data Ethics \& IRB Approval}

Data collection to experimentally validate the efficacy of our approach has been approved by our Institutional Review Board (IRB). It is conducted under a Data Usage Agreement (DUA) with the campus network IT group that restricts and safeguards all WiFi data. To avoid private data leakage, all the MAC IDs and usernames in the syslogs are anonymized using a strong hashing algorithm. The hashing is performed before syslog data is stored on disk under the campus IT manager's guidance, who is the only person aware of the hash key of the algorithm. Any data analysis that results in the users' de-anonymization is strictly prohibited by our IRB agreement and the signed DUA. 

\section{Network Centric Contact Tracing Approach}
\label{sec:sys_arch}
This section presents an overview of our approach, followed by the details of our graph-based contact tracing algorithm.

\begin{figure*} [h]
\includegraphics[scale=0.63]{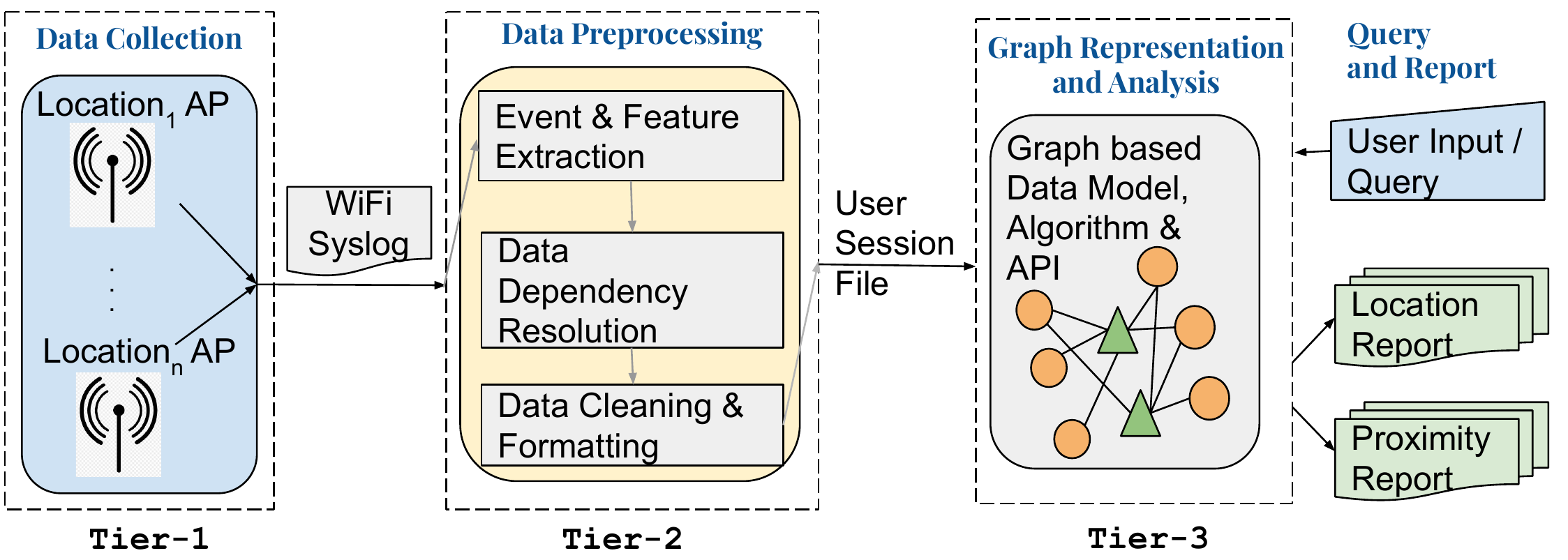}
    \caption{An overview of the tiered architecture used by our approach.}\label{fig:sys_overview}
\end{figure*}

\subsection{System Overview}

Figure \ref{fig:sys_overview} shows that \sysname uses a three-tier pipelined architecture. The data collection tier uses network logging capabilities that are already present in enterprise WiFi systems to collect the {\em WiFi logs of device associations to access points} within the network. Many enterprise IT administrators already collect this data for network monitoring, in which case this already-collected data can also be fed into \sysnames. Otherwise, the IT admins need to turn on WiFi logging to start gathering this data.

The next tier ingests this raw data and converts it into a standard, intermediate format -- i.e., it performs pre-processing of the data. Since the raw log files will have vendor-specific formats, this tier implements vendor-specific pre-processing modules specific to each WiFi manufacturer and its logging format. This tier processes log files in batches every so often and generates data in intermediate form. 

The final tier ingests the intermediate-form data produced by the vendor-specific pre-processor and creates a graph structure that captures the trajectories of all user devices. This tier also exposes a query interface for contact tracing. For each query, it uses the computed trajectories over the query duration to produce (i) a location report listing locations visited by the infected user and (2) a proximity report listing other users in the proximity and their respective durations. As discussed below, this tier uses efficient time-evolving graphs and algorithms to efficiently intersect the trajectories of a large number of devices (typically tens of thousands of users that may be present on a university campus) to produce its report.

\subsection{Basic Contact Tracing Approach}
Consider a WiFi network with N wireless access points that serves M users with D devices. We assume that the N access points are distributed across buildings and other key spaces in an academic or corporate campus and that the location of each access point (e.g., building, floor, room) is known. Large enterprises such as a residential university will install thousands of access points to serve tens of thousands of users. For example, our work uses data from two large universities, one based in the Northeastern USA with 5500 access points and one based in Singapore with 13,000 access points. To manage such a large network, enterprise WiFi networks use controller nodes to administer and manage the APs and the network traffic and provide detailed logging and reporting capabilities.

As a user moves from one location to another, their mobile device (typically a phone) associates with a nearby access point. Since the locations of APs (building, floor, room) are known, the sequence of AP associations over the course of a day reveals the user's trajectory and visited locations. We extract these association events from the WiFi controller logs to reconstruct this trajectory. Typically this information is of the form: {\texttt timestamp, AP MAC address, Device MAC, optional user ID, event-type}, where event-type can be one of association, disassociation, reassociation (when a device wakes up from sleep), un/authorization (the device is not authenticated). Typically when a device switches to a new AP due to user mobility, this is visible to the network in the form of a disassociation event with the previous AP and an association event with a new AP. 

Given this log information, the process of contact tracing a particular user involves two steps: 
\begin{enumerate}
    \item determine all APs visited by the user in the specified time period and
    \item determine all users who were associated with each of those APs concurrently with the infected user.
\end{enumerate}

To do this, we analyze the WiFi logs to construct the time-ordered sequence of sessions that represent the time that the devices spend at each AP. A session is the time period between an association event and a subsequent disassociation event for that AP and device. Since AP locations are known, this sequence of sessions also represents the user's location.

\begin{figure}
\includegraphics[width=0.81\linewidth,keepaspectratio]{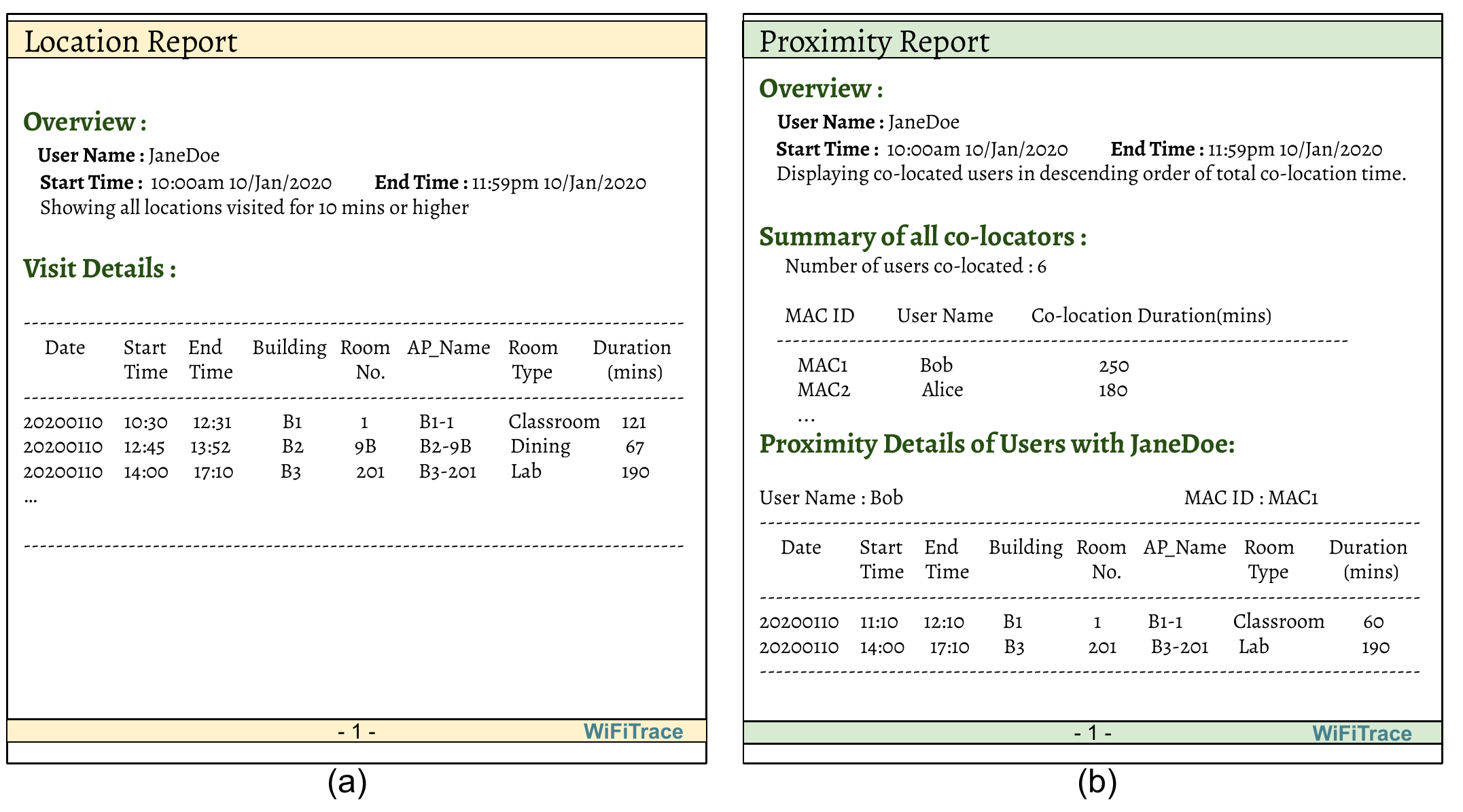}
    \caption{An example contact tracing report produced by \sysnames:
    (a) Patient Report (b) Proximity Report \label{fig:reports}}
\end{figure}



Next, we analyze the log for each AP session to determine all other users \ncz{with} overlapping sessions for the same AP. These are users (i.e., their devices) who were present in the infected user's proximity. Of course, the WiFi log does not reveal the distance between the two users or whether physical contact occurred. Nevertheless, it enables us to determine users at risk by computing the {\em duration} for which the two users were in proximity to one another. In some cases, the location where they were co-located may reveal the degree of risk (e.g., an hour-long meeting in a small conference room or a lecture classroom). To allow health professionals to assess the risk, we generate a location report, showing locations visited and duration of visit by the user, and a proximity report of co-located users at each location and co-location duration. Figure \ref{fig:reports} depicts a sample report resulting from the process.

One practical problem is that an enterprise network with thousands of APs and tens of thousands of devices will generate large log files. For example, one of our campuses' log file contains more than 9 billion events over a 4 month semester period. Thus, scanning the log to compute the location and proximity can be slow and inefficient. To address this, we present an efficient graph-based algorithm based on time-evolving graphs in the next section.

\subsection{Efficient Contact Tracing Using Graphs}
\label{sec:graphs}
To efficiently process contact tracing queries, we model the data as a bipartite graph between devices and APs. Each device in the WiFi log is modeled as a node in the graph;  each AP is similarly modeled as a node. An edge between a device node and an AP node indicates that it was associated with that AP. The time interval annotates each edge ($t_{1}$, $t_{2}$) that denotes the start and end times of the association session between that device and the AP. Note that data is continuously logged to the log files, which causes new edges to be added to the graph as new associations are observed and new nodes to be added as new devices are observed in the logs. Thus, our bipartite graph is time-evolving.

\begin{figure*}
\includegraphics[width=5in]{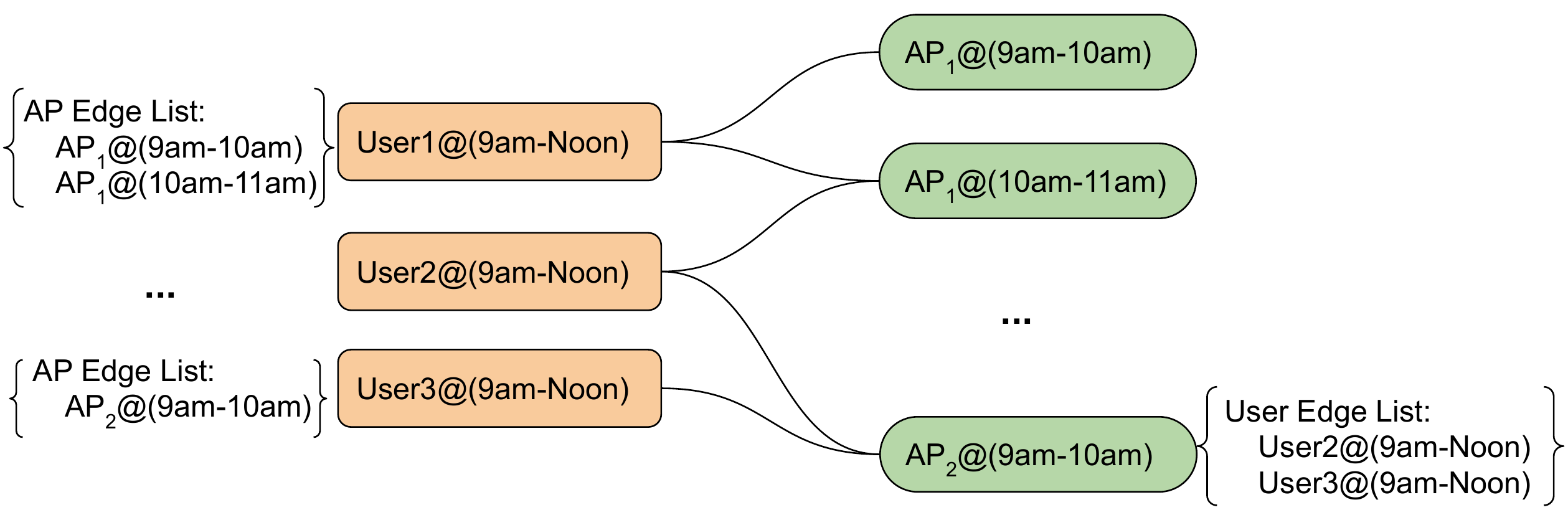}
    \caption{An example bipartite graph shows device to AP association and time-based partitioning of node activity. \label{fig:bipartite}}
\end{figure*}

For computational efficiency, each device and AP node in the graph is limited to a time duration, say an hour or a day.  This is done to limit the number of edges on each node, which can keep growing as the device associates with new APs or APs see a new association session. As a result of associating a time duration with each node, {\em each} device or AP is represented by {\em multiple} nodes in the graph, one for each time duration where there is activity. In this case, we can view the node ID as the mac address concatenated by the time duration. For example,   $MAC_{1}$\@[10:00,10:59], $MAC_{1}$\@[11:00,11:59], represent two nodes for the same device, each capturing AP association edges seen within that period. In the case of AP nodes, this would capture all device association to that AP within those time periods (see Figure \ref{fig:bipartite}).  The duration for partitioning each node's activity in the graph is a configurable parameter. This duration can be chosen independently for a device node and an AP node if needed.

Given such a bipartite graph, a contact tracing request is specified by providing a device MAC address and a duration ($T^{start}$, $T^{end}$) over which a contact trace report should be generated. The query also takes a threshold $\tau$ that specifies only AP sessions of duration longer than $\tau$ should be considered.

The graph algorithm first identifies all device nodes corresponding to this user that lies within the ($T^{start}$, $T^{end}$) interval and identifies all edges from these nodes. These edges represent all AP locations visited by the user, and the session duration represents the time spent at each location. Only edges with the following constraints are considered:  (1) the session must lie within the query time interval, i.e., $[t_1,t_2] \in [{start}$, $T^{end}]$ and (2) the session duration must be at least $\tau$, i.e., $(t_2 - t_1) \geq \tau$.
Edges that do not satisfy either of the above criteria are ignored. The remaining edges are used to enumerate the AP locations visited by the device and the time duration spent at each location.

To compute the proximity report, the algorithm traverses each edge and examines the corresponding AP node. For each AP node, the list of incident edges corresponds to all devices that had active sessions with that AP. The session duration $[t_3$, $t_4]$ on each edge is compared to the infected user's session $[t_1,t_2]$, and the edge is included only of the two-session overlap. This process yields a list of all other users who had an overlapping session with the infected user. The algorithm can also take an optimal parameter $w$ that indicates the minimum overlap in session between the two for the user to be included in the proximity representation, i.e., $ w \geq [t_1,t_2] \cap [t_3,t_4]$. The parameter $w$ specifies the minimum duration of the co-location necessary for a user to be included in the proximity report. Algorithm \ref{algo:algo1} lists the pseudo-code for our graph algorithm. Thus, a time-evolving bipartite graph allows for efficient processing of contact tracing queries over a large dataset.

\begin{algorithm} 
 \caption{Contact Tracing}
 \label{algo:algo1}
 \begin{algorithmic}
     \Procedure{ContactTracing}{$Graph$, $UserID$, $\tau$, $w$, $T^{start}$, $T^{end}$}
     \State devNodesList $\gets$ find all device nodes for UserID in interval [$T^{start}$, $T^{end}$]
     \For{node in devNodesList}
        \State Filter out all edges where session($t_{1}$, $t_{2}$) $\notin$ [$T^{start}$, $T^{end}$] and ($t_{2}$ - $t_{1}$) $<$ $\tau$
        \For{each remaining edge}
            \State Add edge.APLocation and edge Session($t_{1}$, $t_{2}$) to list of locations
            \State Add APnode corresponding edge  to APnodesList
        \EndFor
        \EndFor
    \For{\texttt{node in APnodesList}}
        \State Filter edges where user session($t_{1}$, $t_{2}$) doesn't overlap with colocator session ($t_{3}$, $t_{4}$) \State and $ w \geq [t_1,t_2] \cap [t_3,t_4]$
        \For{each remaining edge}
        \State Add user and device information corresponding to edge list of co-locators.
        \EndFor
     \EndFor
  \EndProcedure
\end{algorithmic}
\end{algorithm}
\section{System Implementation}
\label{sec:system_implementation}

This section presents the implementation of \sysnames, and it is available as open-source code to researchers and organizations who wish to deploy it (source code is available at http://wifitrace.github.io).

\begin{table}
    \begin{center}
    \caption{Graph APIs implemented by our graph-based data representation.}\label{tab:graphapi}

     \begin{tabular}{ l l } \toprule
     Service & Description \\ \midrule
     get\_id & Return Node Id (MAC id for user node/Location id for location node) \\
     add\_neighbor & Add a location neighbor and init edge weight \\
     get\_connections & Return all visited locations \\
     get\_weight(location) & Return edge weight for a location \\
     get\_sessions(user) & List sessions for the user at current location node \\
     get\_name & Return non-time indexed location name\\
     get\_users & Return list of all users\\
     get\_location & Return list of all locations\\
     add\_user\_node & Add a new user node\\
     add\_loc\_node & Add a new location node\\
     add\_edge &  Add a new edge in graph\\
     get\_user\_node & Return user node from graph user node dictionary \\ \bottomrule
     \end{tabular}
 \end{center}
\end{table}

\subsection{Three-tier Architecture}



From the three-tier architecture (see Figure \ref{fig:sys_overview}), our system implementation occurs in the second and third-tier; the first tier is based on the logging capabilities already supported by enterprise-grade WiFi networks, and we currently support WiFi access points from Cisco and HP/Aruba, two large enterprise WiFi equipment vendors. 

\subsubsection*{\textbf{Data Preprocessing}}
We implemented preprocessing code for both Cisco and Aruba networks that takes raw monitoring data and converts it to a standard intermediate data format for our second tier. For HP/Aruba network, \sysname supports the processing of both syslogs (generated by Aruba WiFi controllers) and RTLS logs generated by Aruba APs. Both types of logs provide association and disassociation information. In the case of Aruba RTLS, we log WiFi data directly from the controller nodes using either real-time location services (RTLS) APIs~\cite{ArubaRTLS}. For Aruba syslogs, we periodically copy the raw syslogs generated by the controller and preprocess this raw data. Finally, for Cisco networks, we log WiFi data directly from the network using the Cisco Connected Devices (CMX) Location API v3~\cite{CiscoCMX}. All of these preprocessor scripts convert raw logs into the following standard record format:
\begin{verbatim}
    Timestamp, AP Name or ID, Device MAC ID, event type, (optional) Username
\end{verbatim}

By default, we assume anonymized (or hashed) device MACs and usernames. We also assume a separate secure file containing a mapping of real names to hashes. 
\apt{While the association, disassociation, reassociation, and drift messages from the syslog give us spatio-temporal details about the various devices on the network, the authorization events provide us with details about MAC ID and username, aiding us to create a device-user mapping. The device-user mapping helps us count each user once by considering only the highly mobile user device among multiple user-owned devices. The identified highly mobile device is most commonly a smartphone because it gets carried around by users everywhere.} 

\subsubsection*{\textbf{Graph Representation and Analysis}}
Our third tier supports contact trace querying. A query is of the form (hashed) username or device MAC, start duration, end duration, threshold $\tau$, and co-locator threshold $w$. Internally the data generated by the pre-processing code is represented as a bipartite graph, as discussed in Section \ref{sec:graphs}. Our system supports various queries on this graph through a graph API depicted in Table \ref{tab:graphapi}. This graph API is used to implement the graph algorithm described in Section \ref{sec:graphs}. 

The algorithm yields a {\em location report}, which shows all locations (APs) visited by the user for longer than $\tau$ and a {\em proximity report} that shows all users who were connected to those APs for a duration greater than $w$. Figure \ref{fig:reports} shows a sample location and proximity report generated by our system. 

In addition to human-readable query reports, our system can optionally output query results in JSON format, convenient for visualization or subsequent processing. Our system also supports additional report types beyond location and proximity reports. For example, it can produce reports of additional users who visit a location {\em after} the infected user has departed from that location. This is useful when a location has high-contact surfaces that may transmit a contagious disease even after an infected user departs. Such a report can be produced by specifying a window parameter that specifies the time other users are identified as at-risk at each location after they depart.

\section{Deployment and Validation}


In this section, we describe our deployment and validation results for contact tracing.

\subsection{Deployment}
We deployed \sysname on a university campus in the northeastern USA and one in Singapore. Both campuses have large WiFi networks, one with 5500 HP/Aruba APs and a mixed Cisco/Aruba network of 13,000 APs. While we had originally developed this tool to fight the outbreak of meningitis on our campus, university health officials from both campuses viewed our solution as a scalable contact tracing method for COVID-19.

Even though \sysname has been operational for several months now,  as of May 2020, \final{neither campuses had} experienced a situation that requires using \sysnames. This was due to strict lockdown measures with both residential universities switching to online learning. Students were instructed to vacate their dorms, and work-from-home policies were enforced. Except for a small number of students unable to return to their home countries (due to global lockdown), both campuses have been largely empty until recently (Sep 2020 onwards) when the Singapore campus started letting more students on-site. One of our campuses saw a single case of an employee at a high-risk of COVID-19. However, manual contact tracing had determined the case as a low threat to others. As such, university health professionals did not need to perform further contact tracing. 

\subsection{Validation}

We conducted a small-scale user study to gather ground truth data to validate four questions related to the use of passive WiFi sensing for contact tracing: 
\begin{itemize}
    \item V1: How accurate are WiFi access point associations at revealing the true user location? 
    \item V2: How accurate are WiFi session durations at revealing the true duration a user spent at a location? 
    \item V3: Do co-located WiFi device sessions accurately show co-located users at those locations?
    \item V4: \apt{How do AP density, phone activity, user movement type, and phone make/model/OS impact device registration events seen by the WiFi network? In particular, do these factors result in a temporal lag in the extracted trajectories?}
\end{itemize}


\subsubsection*{Dataset} We \apt{collected data from a group of eighteen volunteers over a period of 37 days with each volunteer walking to various locations, and spending variable lengths of time at each location, around our campus} while carrying their mobile devices. Each user manually logged the entry and exit times at each location and the paths they took from one location to the next. For each user, we computed a location report containing all visited locations (assuming a threshold $\tau=0$) and compared the locations as seen by the WiFi network to the ground truth locations recorded by each user. Some of the user's devices' trajectories were correlated, which meant the users were co-located whenever (their devices were) connected to the same AP concurrently. Overall, about 19,000 unique locations were visited by our volunteers. 

\subsubsection*{Results}
\begin{figure}[t]
    {\begin{tabular}{cc}
    \includegraphics[width=2.7in,keepaspectratio]{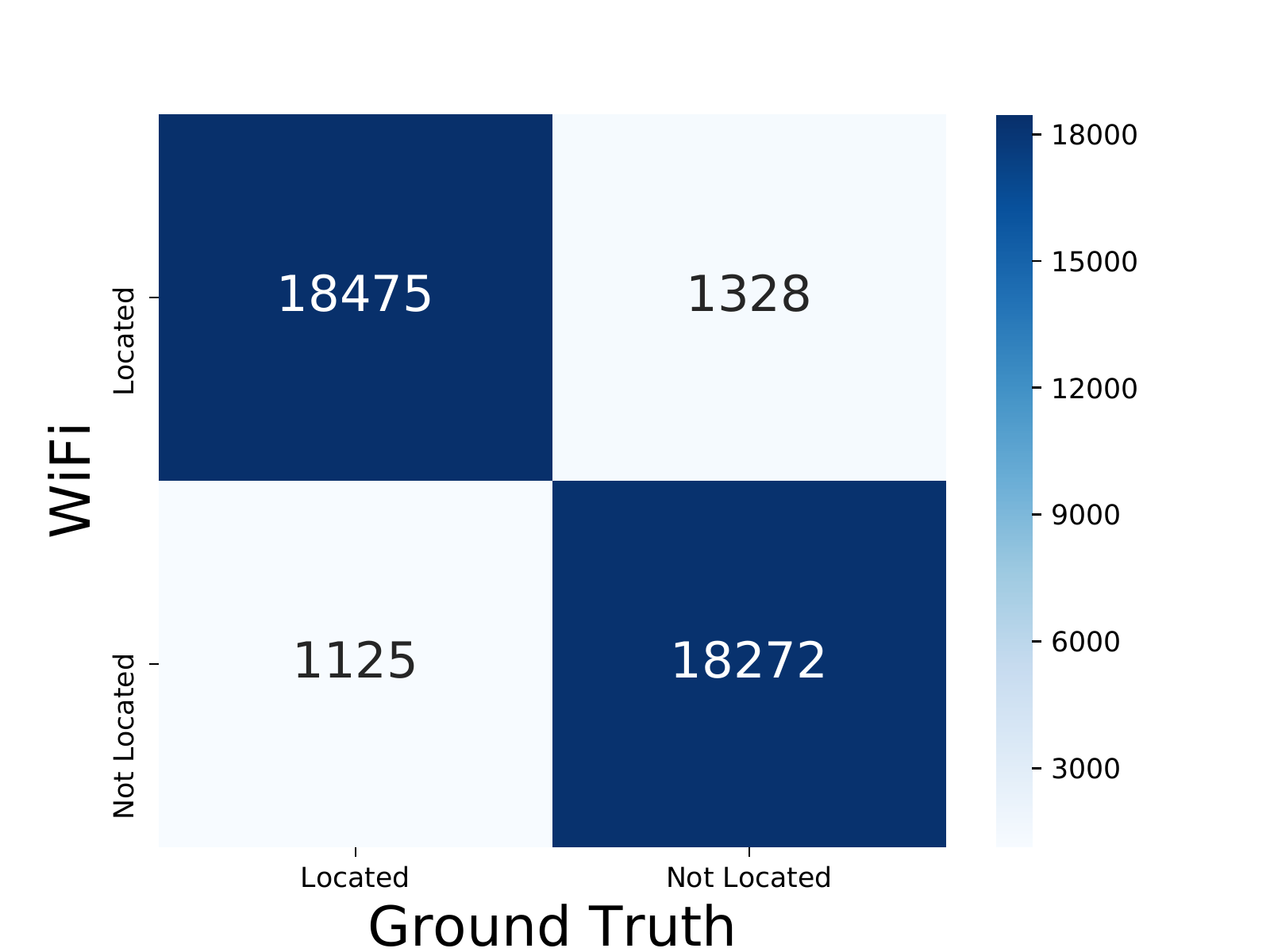} &
    \includegraphics[width=2.7in,keepaspectratio]{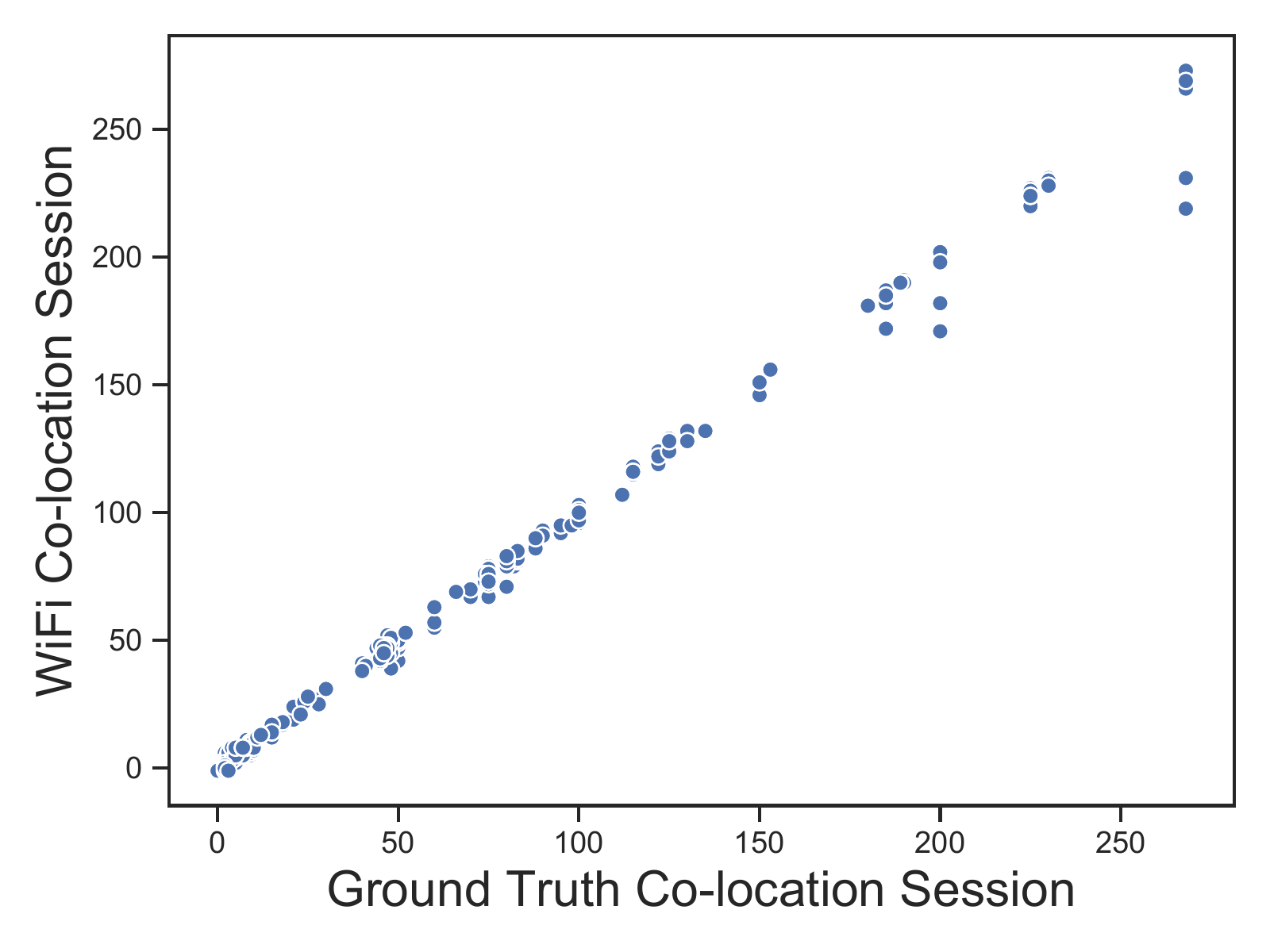}\\
    (a) & (b)
    \end{tabular}
    }
	\caption{(a) Confusion Matrix (b) Scatter Plot displaying ground truth and WiFi based session duration \label{fig:confusion_matrix}}
    \vspace{-0.3cm}
\end{figure}

\apt{To answer V1, we compared the locations reported by WiFi with the actual ground truth location recorded manually by each volunteer.  Figure~\ref{fig:confusion_matrix}(a) shows the confusion matrix for this comparison. Specifically,  an overlap in WiFi and ground truth \textbf{\textit{located}} labels indicate the inferred location from a WiFi AP of a user matches their logged location -- true positive. An agreement in WiFi and ground truth \textbf{\textit{non-located}} labels imply a user was, in fact, not physically situated where the WiFi AP had not inferred them to be -- true negative. Similarly, two mismatch possibilities can occur. The first is the WiFi had missed inferring the user location to where a user was actually localized (WiFi \textbf{\textit{non-located}} and ground truth \textbf{\textit{located}}) -- false negative. The second is that WiFi had inferred the user to be situated in a spot the user was not in (WiFi \textbf{\textit{located}} and ground truth \textbf{\textit{non-located}}) -- false positive. Both errors could arise due to temporal delay in AP hopping or skipping, where the user's device had maintained a previous AP connection, despite the user already transitioning to a different spot.}

Overall, \sysname yields a precision of 0.93, a recall of 0.94, and a high F1-score of 0.93. In other words, the inferred location matches the ground truth location with high accuracy. \apt{A deeper analysis of the errors shows that they mainly occur when a user transitions between different locations quickly -- on the order of tens of seconds. Even in these cases, the true location of these quickly moving users is usually just off by one AP.} Figure~\ref{fig:confusion_matrix}(b) shows a scatter plot of the session durations reported by \sysname compared to the ground truth and demonstrates a good match between \sysname and the ground truth. The small errors in the figure occur when devices enter or exit the area as these devices will need additional time to switch networks.

\begin{figure}[h]
\includegraphics[width=3.2in,keepaspectratio]{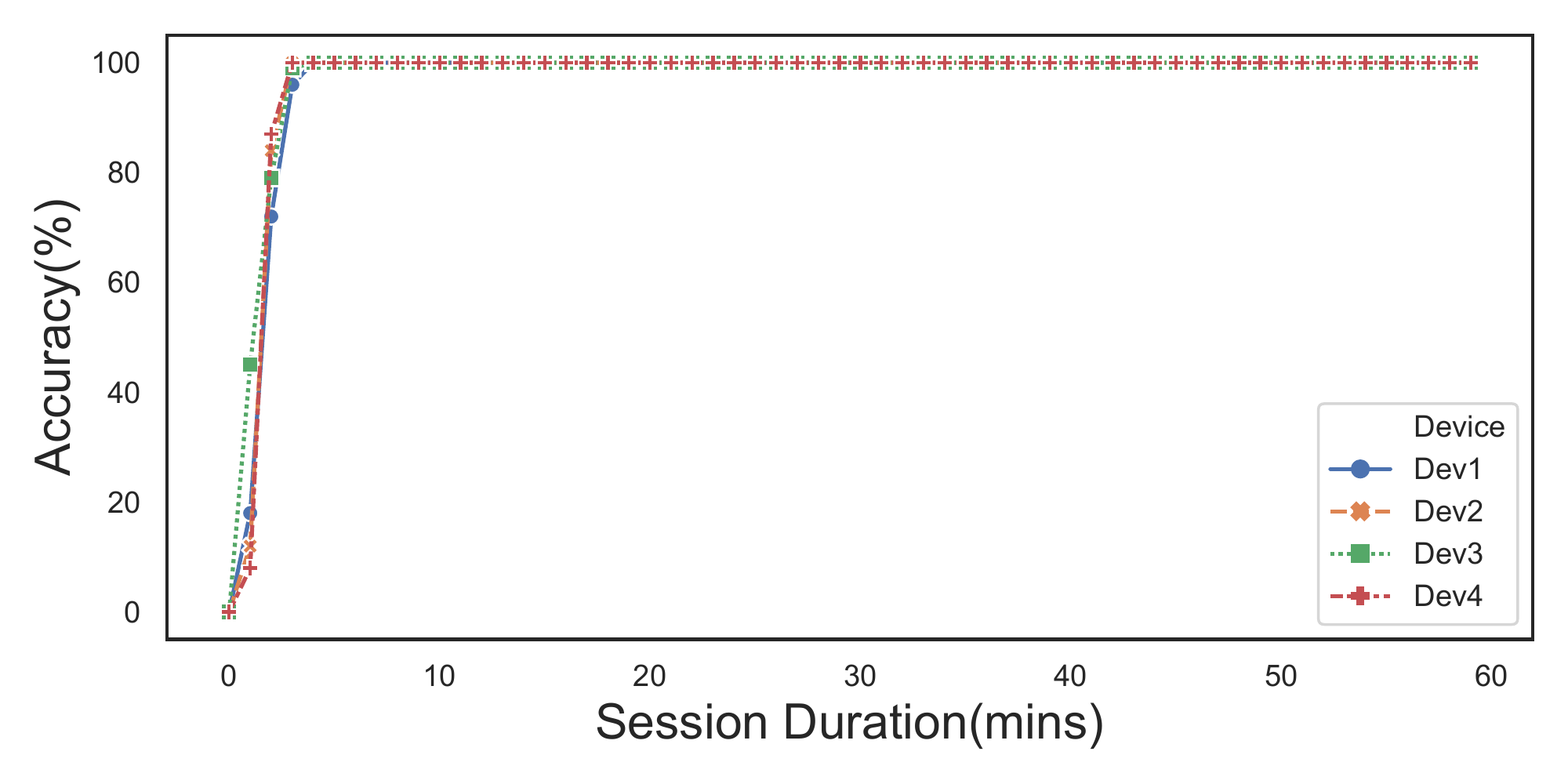}
    \caption{Accuracy of inferring user locations for varying WiFi session duration.}
    \label{fig:loc_accuracy}
\end{figure}

To answer V2, Figure~\ref{fig:loc_accuracy} plots the inferred location's accuracy for varying session lengths observed across four different mobile devices made by Apple, Samsung, Motorola, and LG. \apt{Our tool yields 100\% accuracy whenever a session length exceeds 3 minutes, which is sufficient accuracy for a contact tracing application.} 

Next, to answer V3, we validated the accuracy of co-locations by using \sysname to generate the proximity report for each device and comparing it to the ground truth trajectories reported for each device. \sysname can capture co-located devices (and users) with high accuracy for sessions exceeding 3 minutes, as shown in Figure~\ref{fig:loc_accuracy}. 

However, as noted earlier, if a user moves fast (i.e., there are short transitions), the locations can be off by one AP cell. This implies that the network can see two fast-moving devices near one another connected to adjacent APs, rather than the same one.  Fortunately, this does not hamper the efficacy of contact tracing as two users need to be near one another for a period of time (e.g., 15 minutes or more) to be considered at-risk. \sysname accurately capture these longer co-located times.

\begin{figure}[h]
\includegraphics[height=1.5in,keepaspectratio]{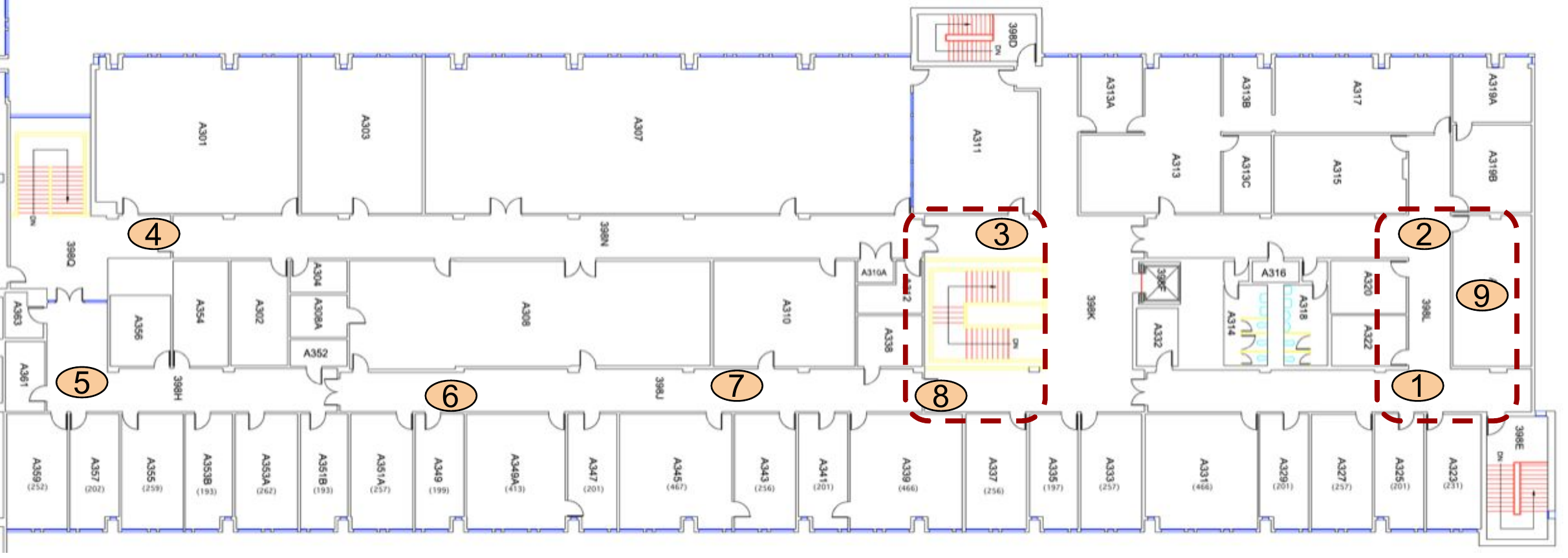}
    \caption{\apt{Floor Map with AP locations  \label{fig:floor_map}}}
\end{figure}

\begin{figure}[h]
\includegraphics[width=5.5in,keepaspectratio]{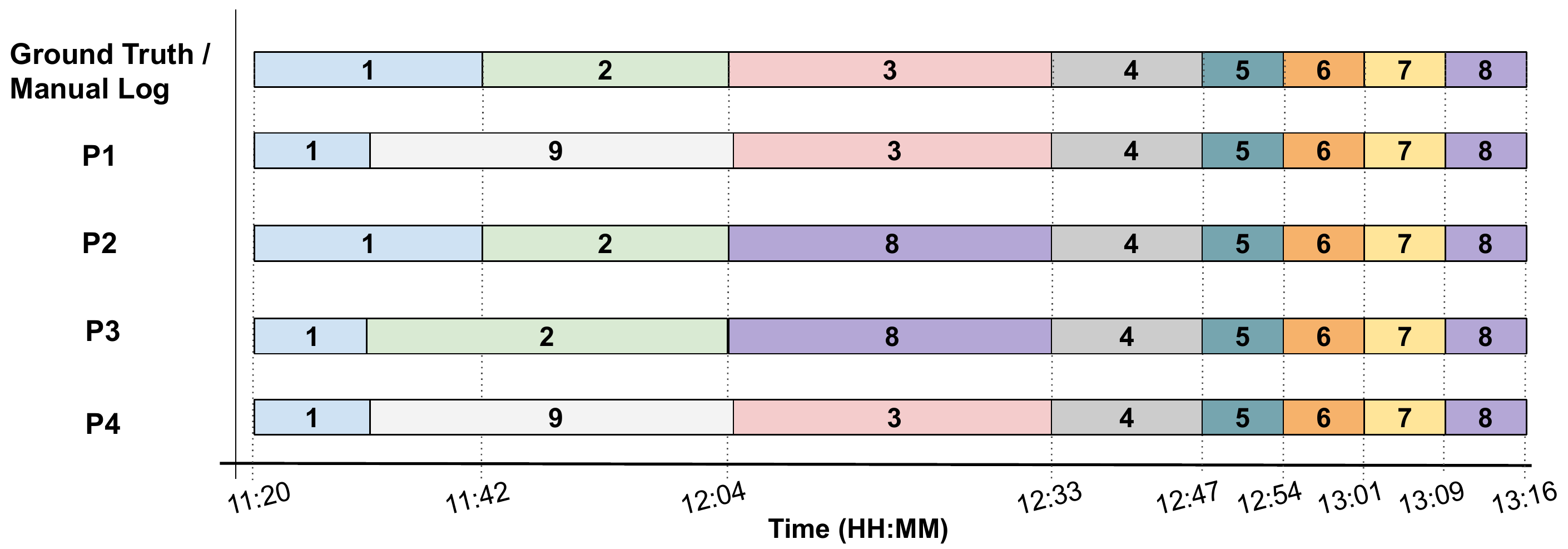}
    \caption{\apt{Temporal lag between computed WiFi Trajectories of active Android Phones of different OS versions as compared to Ground Truth User Trajectory.}}
     \label{fig:dev_traj}
\end{figure}

\apt{Finally, to answer V4, we measured the time lag in WiFi logs extracted against ground truth manual trajectory logs. This experiment used four different smartphones (Apple, Samsung, Motorola, and LG) that were streaming a YouTube video continuously (i.e., they were always active on the WiFi network) and traveling through the same paths and floors. Figure~\ref{fig:floor_map} shows the single floor route covered by 9 APs. The path chosen moved from AP1 to AP2, AP3, etc., and eventually circled back to AP1. AP9 was never exclusively visited, as shown in the ground truth trajectory in Figure \ref{fig:dev_traj}. The floor was intentionally selected for having both a sparse and dense AP dispersion. We repeated this experiment using 5 different transition styles: walking casually, hustling through several locations, and being in stationary spots between 1, 5, and 20 minutes.}


\apt{Figure~\ref{fig:dev_traj} shows the trajectory connections across APs for four devices, P1 to P4, that moved from AP1 through A9 and were stationary at each AP for 5 to 20 minutes. For example, as a user with device P1 moved from AP1 through AP8 along the same path as indicated by the ground truth trajectory. However, device P1 was first connected to AP1 for a short duration and then connected to AP9 even though the user moved to AP 2. P1 then connected correctly to AP 3, 4, 5, etc., as the user moved across the path. In contrast, P2 was connected to AP1 and 2 but connected to AP8 instead of AP3, even though it followed the same path as P1. We believe this was because AP3 and AP8 were on opposite sides of a large open area, and thus the signal from AP8 was still strong even near AP3. P3 and P4 encountered similar AP connectivity with AP2/9 and AP3/8, respectively.} 


\apt{To rectify the errors due to devices connecting to nearby APs in open areas or areas where there are many APs, we can create zones as indicated by the boxes in Figure ~\ref{fig:floor_map}. For example, APs 1, 2, and 9 form a zone Z1 while APs 3 and 8 form another zone Z2. This grouping of APs into logical zones helps eliminate the errors shown above caused by devices connecting to nearby APs instead of the closest AP.}

\begin{table}[ht]
\begin{minipage}[c]{0.45\linewidth}
     \centering
    \includegraphics[width=0.99\columnwidth]{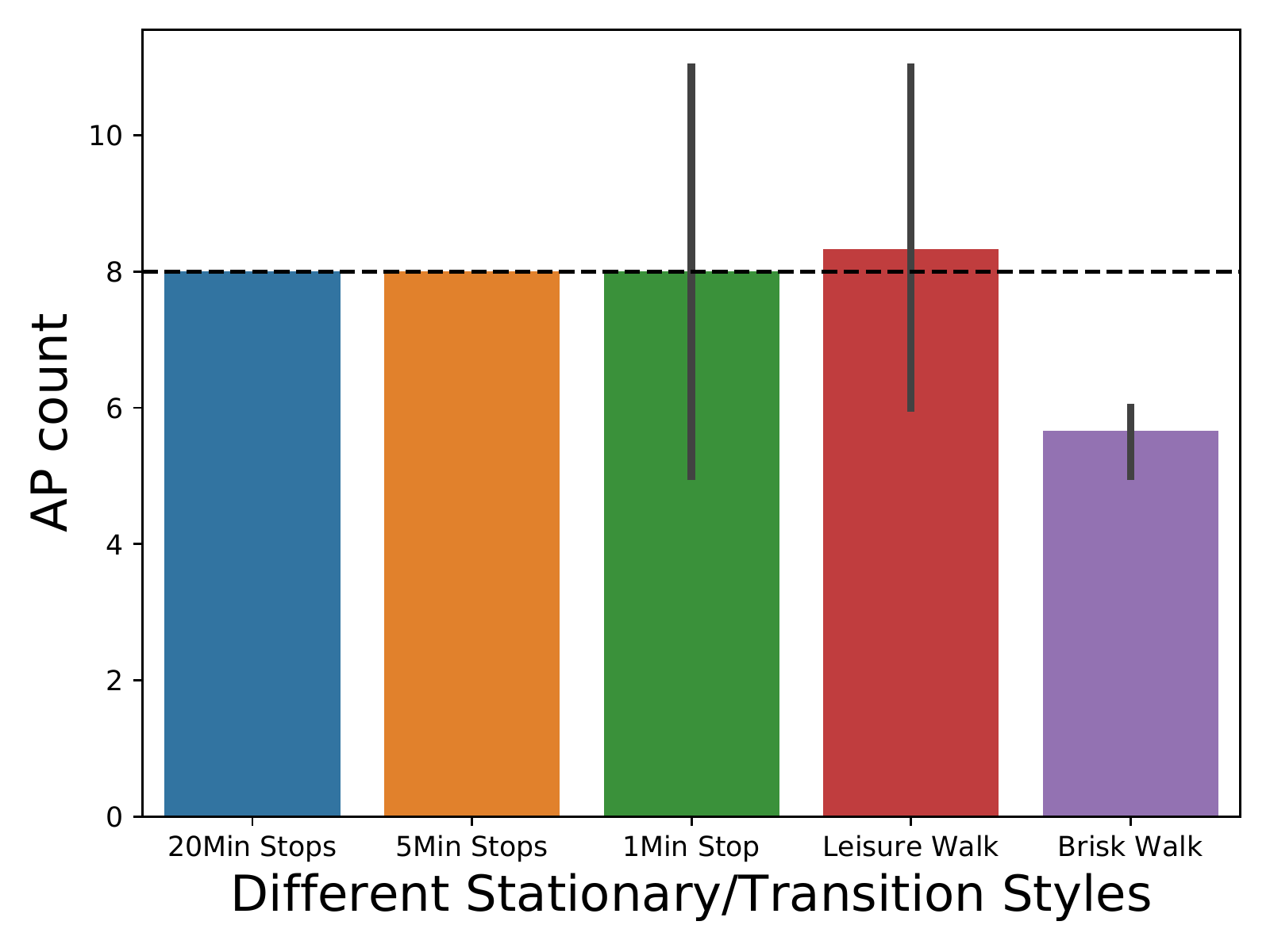}
    \captionof{figure}{\apt{Number of AP Hops encountered for various trajectory styles with varying stop durations for different models, make and OS versions of Android Mobile Phones  \label{fig:ap_hops}}}
\end{minipage}
\hfill
\begin{minipage}[c]{0.45\linewidth}
\centering
  \includegraphics[width=0.99\columnwidth]{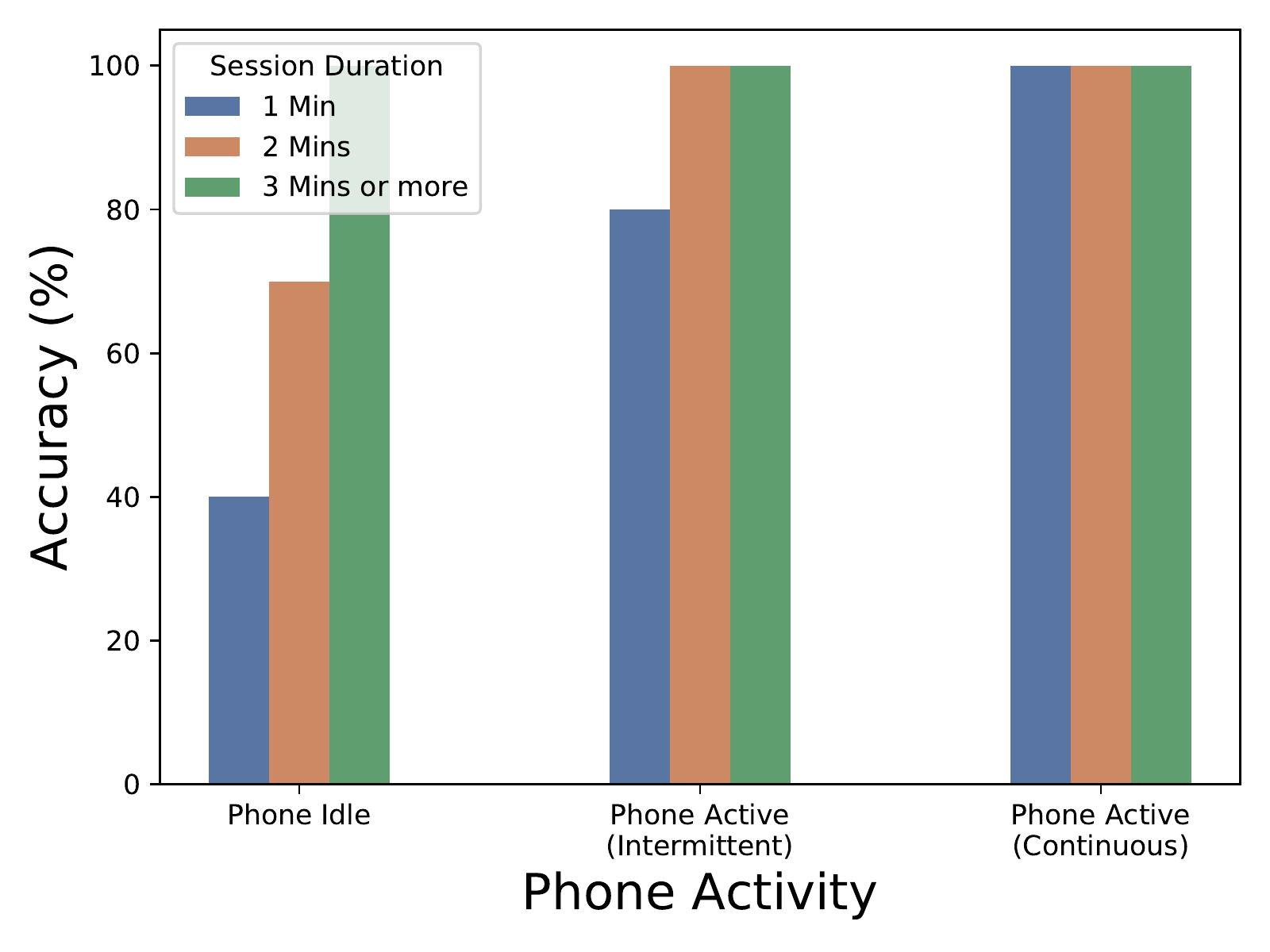}
    \captionof{figure}{\apt{WiFi location Sensitivity for various phone activity scenarios across various session durations as computed across various phone models, makes, and OS (iOS and Android)}}
     \label{fig:sensitivity}
 \end{minipage}%
\end{table}

\apt{Figure~\ref{fig:ap_hops} plots the number of AP hops due to different transition styles. Overall, each device experienced 8 AP hops when moving from AP1 back through AP8 using different transition styles. We observed more variance in AP \ncz{hopped} when a user was moving continuously -- i.e., without making stops. For example, a user walking leisurely through the planned path recorded between 6-11 AP hops, while a brisk walk recorded 5 AP hops on average. This result shows that when no/short stationary stops are made, there is a higher tendency for AP scans to increase.}

\begin{table}[ht]
\begin{minipage}[c]{0.48\linewidth}
     \centering
    \includegraphics[width=0.99\columnwidth]{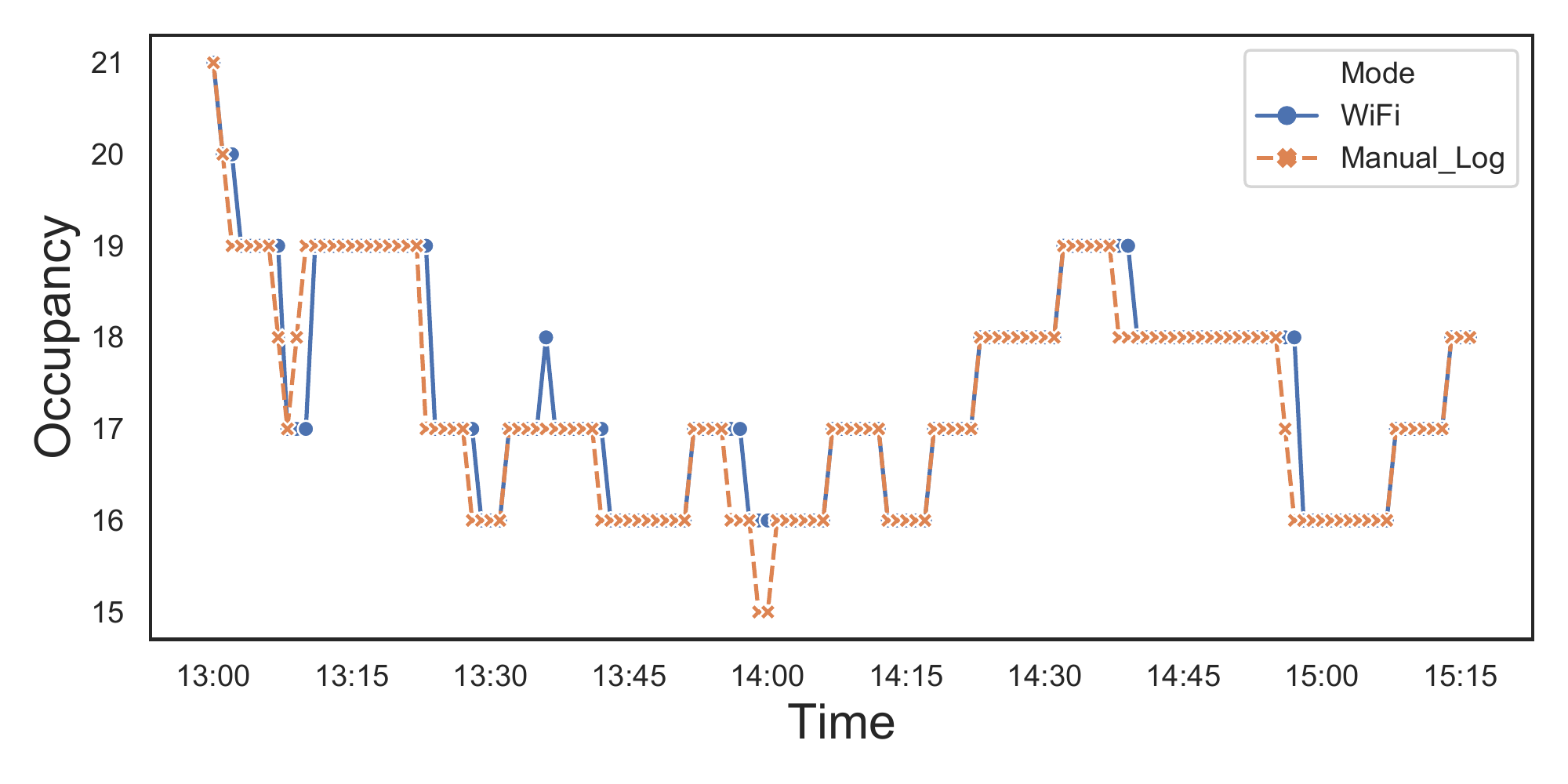}
    \captionof{figure}{Comparison of the number of users co-located at an access point with the ground truth.}
     \label{fig:occupancy}
\end{minipage}
\hfill
\begin{minipage}[c]{0.48\linewidth}
\centering
\caption{Dataset Characteristics}
   \begin{tabular}{llccc} \toprule
        Item  & US University & Singapore University \\ \midrule
       Users  & $\approx$ 38000 & $\approx$ 50000 \\
        AP & $\approx$ 5500  & $\approx$ 13000 \\
        Buildings & 230 & 240 \\
        Time Span & Jan-May 2020 & Feb-May 2020 \\ \bottomrule
    \end{tabular}
    \label{tab:dataset}
    \vspace{-0.6cm}
 \end{minipage}%
\end{table}

\apt{We measured the accuracy of the spatio-temporal values from the WiFi logs when phones were engaging in different activities such as idling, intermittent use (e.g., occasionally responding to messages), and continuous use (e.g., video streaming). Figure~\ref{fig:sensitivity} shows that the accuracy when the phone was idling was the worst, while continuous phone usage had the best accuracy. Specifically, we find that the temporal accuracy values for phone idling are approximately 40\% for sessions less than a minute, 75\% for sessions between 1-2 minutes, and 100\% for session durations more than 3 minutes. On the other hand, continuous phone use guarantees device associations with AP changes indicating that the user's presence will be recorded. User sessions less than a minute for intermittent use maintains an accuracy of at least 80\%. In the most practical use-case, we expect that users transitioning around campus would be engaged in intermittent phone use as they periodically reply to messages or emails. In the worst case, \sysname still maintains high accuracy even when the phone is idling for at least 3 minutes long stationary periods at a location. Thus, \sysname is still useful as a contact tracing application for use cases, such as COVID-19, where individuals need to spend at least a 15-minute overlap time over a 24-hr window period with an infected person ~\cite{cdcDigitalTool2} for the infection to spread.}

Finally, we count the number of users entering and leaving library rooms and compare it to the number of devices (users) reported by our tool.  As shown in Figure~\ref{fig:occupancy}, the automatic count generated by \sysname closely mimics the ground truth. The small mismatches occur due to short WiFi sessions (implying a user is present only for a brief period or when their devices did not switch to a new AP from a previous one). The user counts remain accurate for all sessions that exceed a few minutes as their devices will eventually switch to the closest AP. 

\apt{Overall, our user scale study shows that \sysname can answer questions V1 to V4 with sufficient accuracy to support its use as a contact tracing application.}

\section{Experimental Evaluation}
\label{sec:evaluations}
In this section, we describe case studies that evaluate our contact tracing tool and our graph algorithms. Further, we discuss the limitations of utilizing WiFi sensing.

\subsection{Dataset and Methodology}

\apt{To validate \sysname, we used the production WiFi logs from our university WiFi network.} 
Table~\ref{tab:dataset} summarizes the characteristics of the WiFi logs. Our US university uses an Aruba network of 5,500 APs deployed across 230 buildings. It has approximately 38,000 users comprising 30,000 students and 8,000 faculty\/staff. The dataset spans between Jan 2020 to May 2020, including a full campus-wide COVID-19 lockdown during Spring break (mid-March). 

\begin{figure} [h]
\includegraphics[width=2.1in,keepaspectratio]{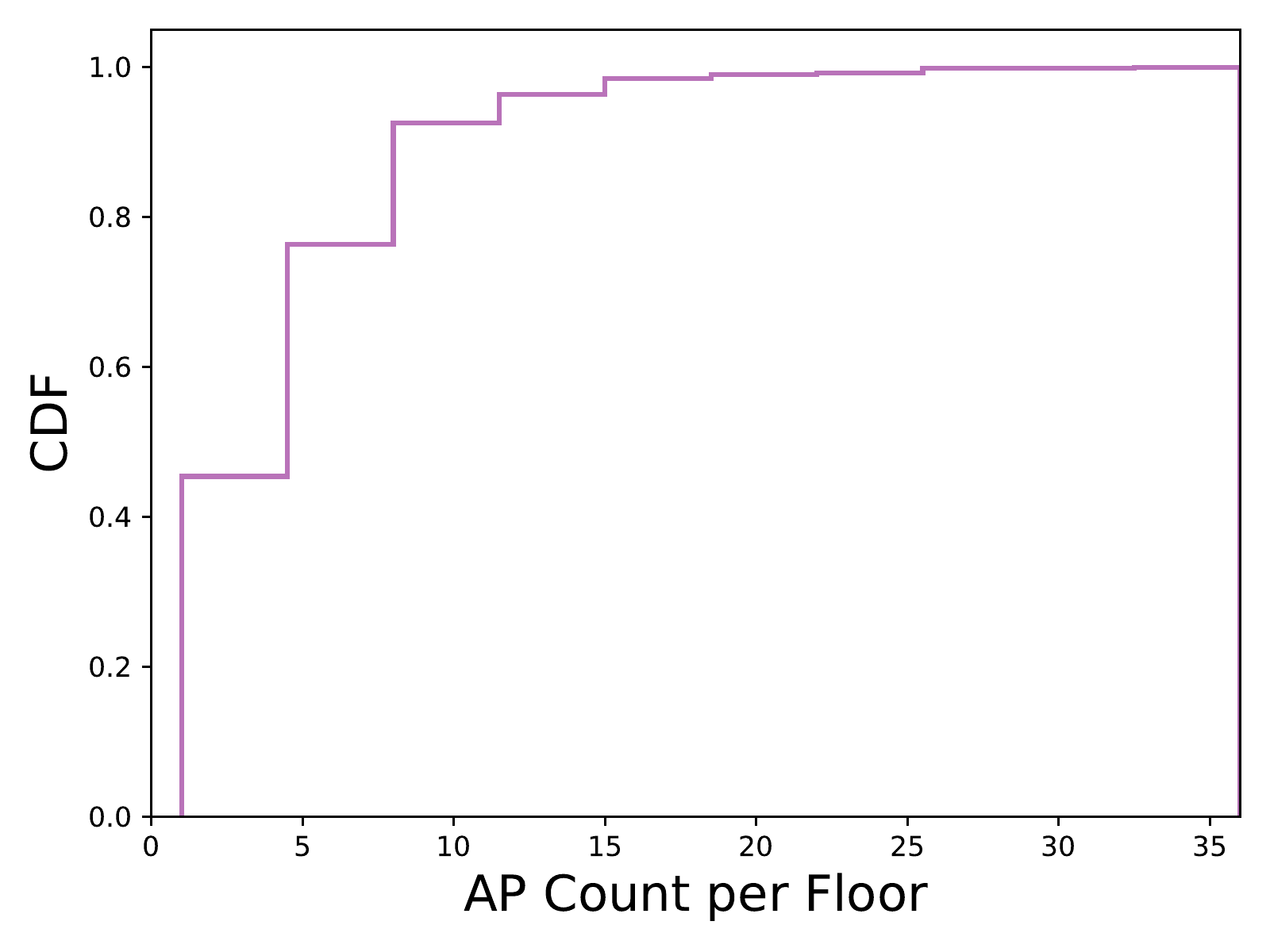}
    \caption{Distribution of APs per floor  \label{fig:ap_dist}}
\end{figure}

\apt{The APs are spread across the US campus, with a mean of ~23 APs per building. We found that 90\% of the buildings have up to 45 APs, and few tall buildings (dorms and library) with a high number of floors have more than 100 APs installed inside. As shown in Figure~\ref{fig:ap_dist} an average of 6 APs is installed per floor. }

We also collected data from a Singapore university that has a mixed Aruba and Cisco network comprising 13,000 APs deployed across 240 buildings. It has approximately 50,000 users comprising 40,000 students and 10,000 faculty/staff. The Singapore dataset spans between Feb 2020 to May 2020. \ncz{It includes} a COVID-19 quarantine plan, which was progressively introduced by the Singaporean government, ending with a full lockdown similar to the US University. 


\subsection{Case Study}

\subsubsection{Efficacy of Contact Tracing}
Since a real disease outbreak is yet to occur on either campus, we emulate how our tool works under emulated diseases. We pick three diseases, each with a different incubation period, which would require contact tracing for a different number of days.
\begin{itemize}
    \item Seasonal Influenza: Incubation period $\approx$ 1-3 days; contact tracing 2 days
    \item COVID-19: Incubation period $\approx$ 4 days, contact tracing 4 days
    \item Measles: Incubation period $\approx$ 8-12 days, contact tracing 10 days
\end{itemize}

We randomly choose a user from our dataset and assume they are infected with one of the above diseases and use our tool to compute the number of locations visited by the user over that period and the number of co-located users. We perform contact tracing assuming $\tau$ = $\omega$ = 10 mins and $\tau$ = $\omega$ = 30 mins, which implies location visited for at least 10 (or 30) minutes and co-location of at least 10 (or 30) minutes. For each disease, we repeat each contact tracing experiment for 50 randomly selected students and then 50 randomly chosen faculty or staff users.

\begin{figure}
    {\begin{tabular}{cc}
    \includegraphics[width=2.7in,keepaspectratio]{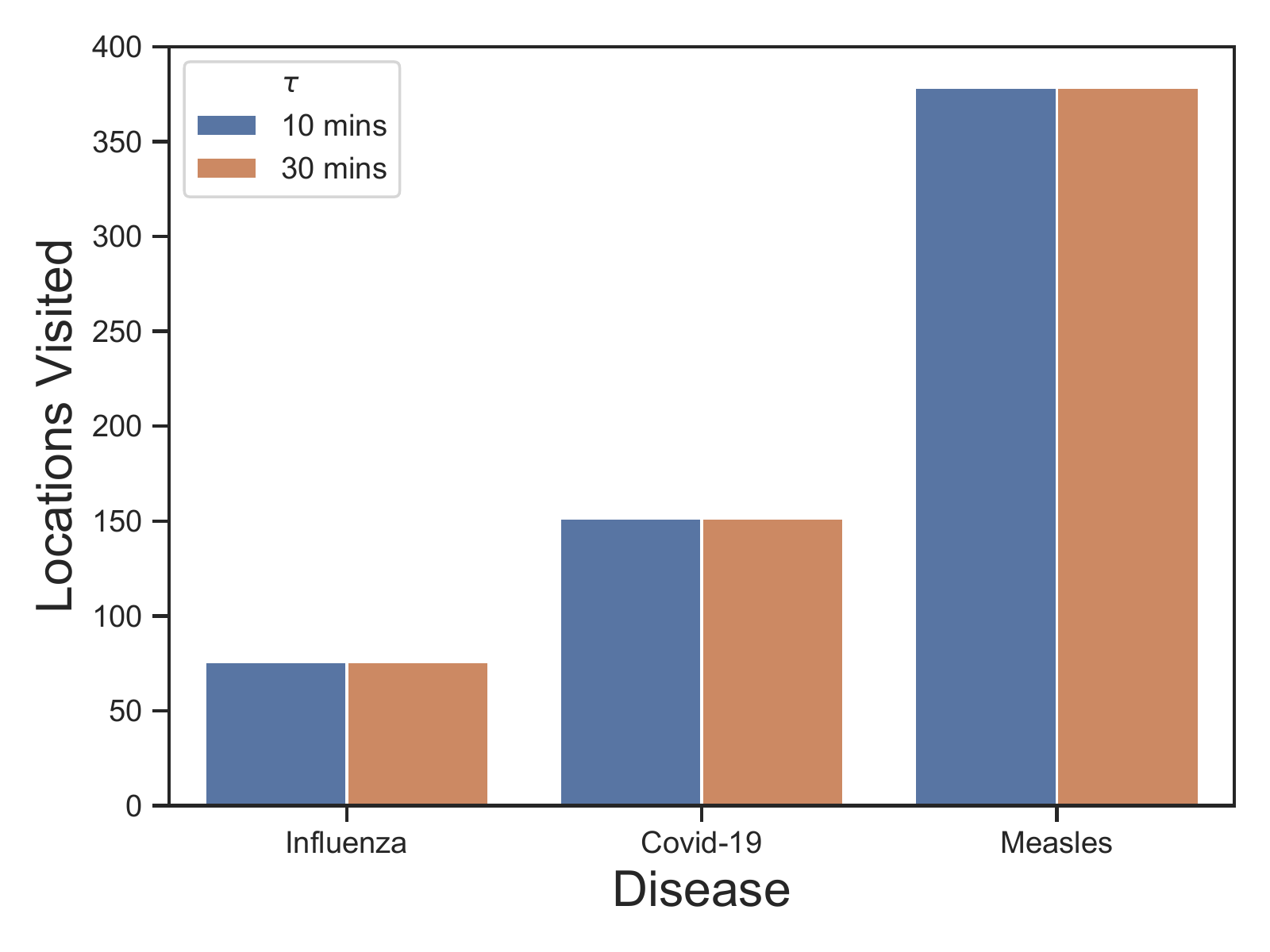} &
    \includegraphics[width=2.7in,keepaspectratio]{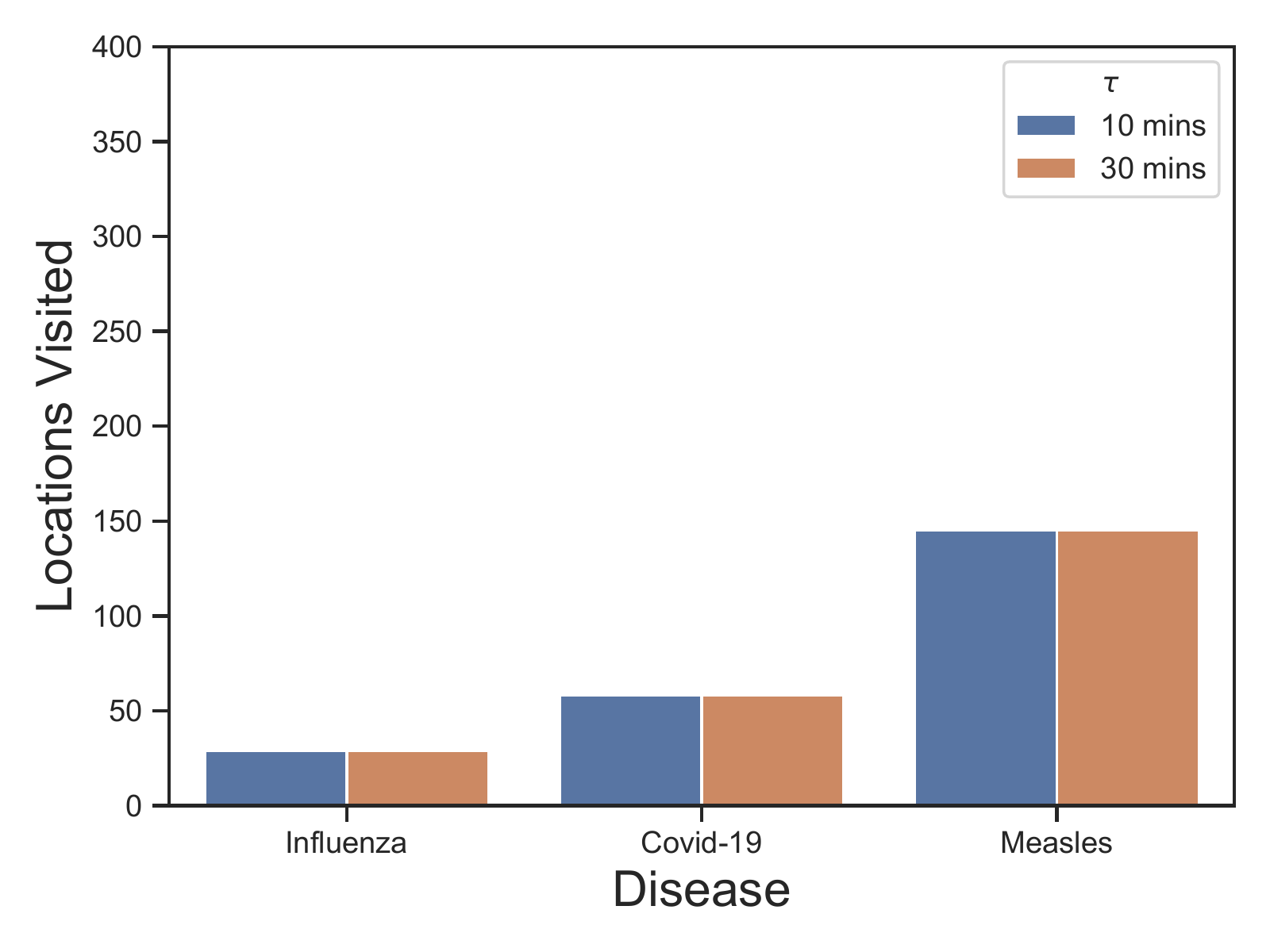}\\
         (a) & (b)
    \end{tabular}
    }
    \caption{Cumulative location count for various diseases for $\tau$ (10mins, and 30mins) for (a) Student User (b) Non-Teaching Faculty Staff \label{fig:loc_faculty}}
    \vspace{-0.3cm}
\end{figure}

\begin{figure}
    {\begin{tabular}{cc}
    \includegraphics[width=2.7in,keepaspectratio]{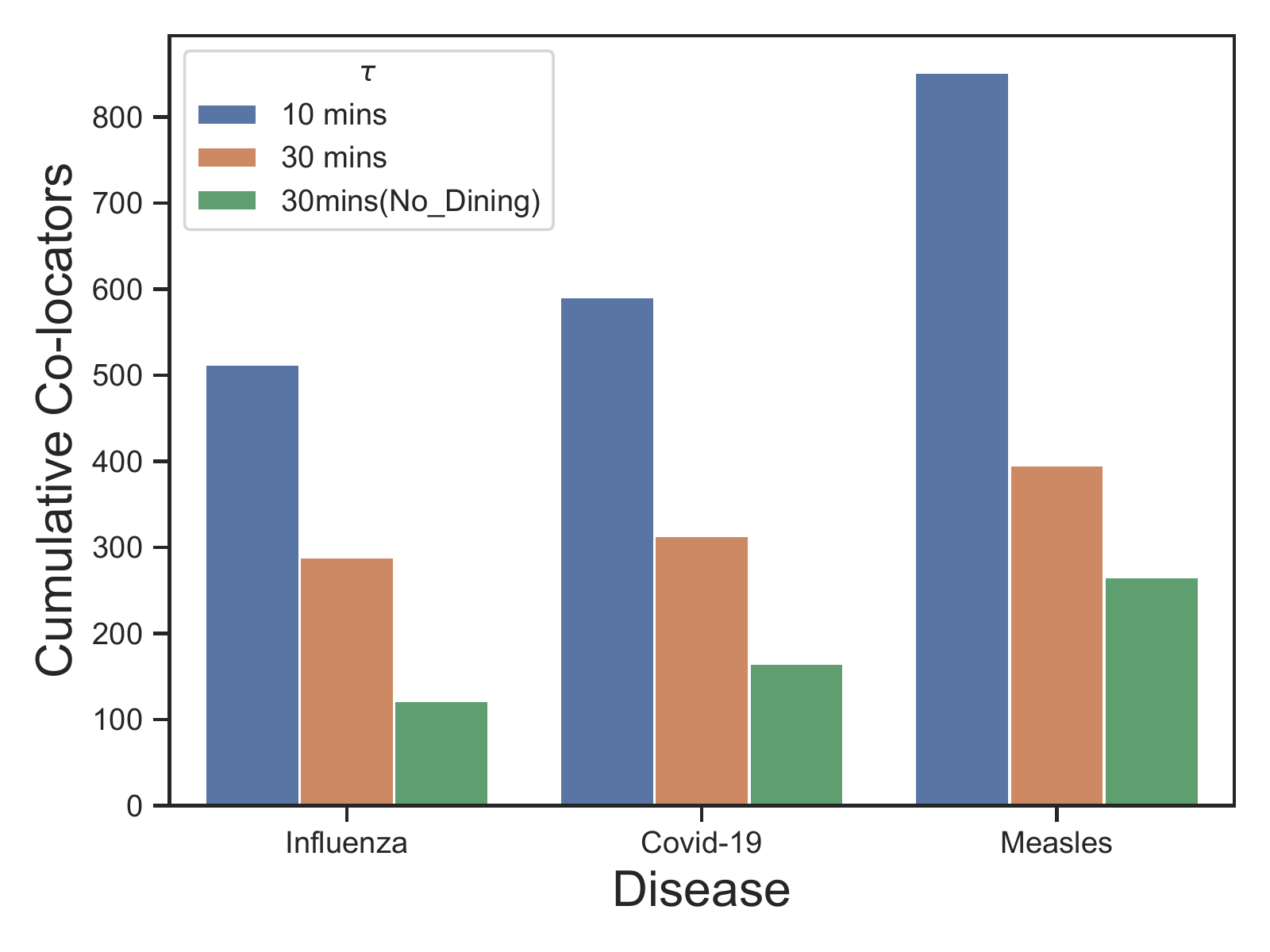} &
    \includegraphics[width=2.7in,keepaspectratio]{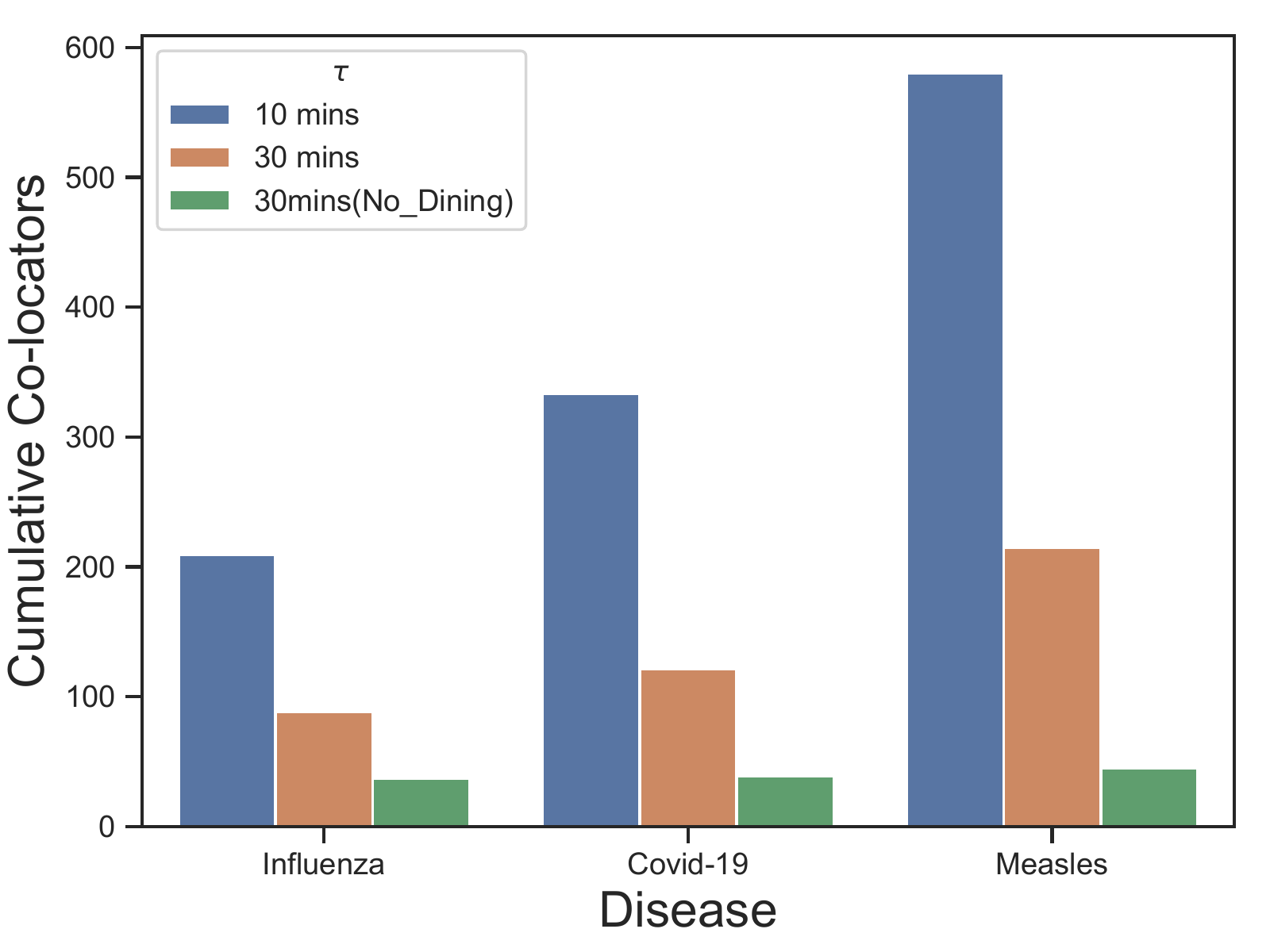}\\
         (a) & (b)
    \end{tabular}
    }
    \caption{Cumulative Co-locator count for various diseases for $\tau$ (10mins, 30mins and 30mins excluding dining) for (a) Student User (b) Non-Teaching Faculty Staff \label{fig:bar_faculty}}
    \vspace{-0.3cm}
\end{figure}

Figure \ref{fig:loc_faculty} shows how the number of locations visited by an infected user grows as the duration of contact tracing increases between 2 - 10 days (for Influenza and Measles, respectively). We find that the number of location visits is insensitive to $\tau$ beyond $\tau$ $>$ 10 mins (we discuss this in detail in the next section). A student visits $\approx$ 37 locations, while a faculty/staff user is less likely to transition around campus and visits $\approx$ 15 locations per day.

Figure \ref{fig:bar_faculty} depicts the proximity results from our contact tracing experiment. With $\tau$ = 10 min, the system yields many co-located users, specifically 500 co-located users for Flu over a 2-day period and over 800 users for Measles over a 10-day period for a student. With $\tau$ = 30 min, the number of co-located users decreases slightly to 300 users for flu and 400 users for Measles. In contrast, the co-location count is lower for staff users, when $\tau$ = 10 mins, but projects to affect about 200 users (during a Flu) and 500 users (during a Measles outbreak). $\tau$ = 30 mins will substantially lower the numbers to 100 and 200, respectively. These results yield the following insights:
\begin{enumerate}
    \item We note that the number of co-locators does not increase linearly with an increase in contact trace duration. The growth is \emph{sublinear}, indicating that users have a limited social circle (of users), and these interactions last over several days.
    \item \apt{It is infeasible to contact trace several hundred users for each infected user manually. \sysname can address this problem by setting the parameters $\tau$ and $\omega$ carefully, balancing the duration of locations and co-locations based on the disease at hand. For example,  $\tau$ = 10 min could result in a high rate of chance co-location. Choosing $\tau$ = 30 mins and $\omega$ be 15 or 30 mins may yield better results. The tool subsequently outputs a manageable number of cases for manual contact tracing investigation. }
\end{enumerate}

Further, our results show that common areas such as cafeterias substantially increase the co-location counts. Accordingly, it is easier to filter out these AP sessions to determine users with higher risk. As shown in Figure \ref{fig:bar_faculty} (a) and (b), the number of co-locators drops considerably once cafeteria visits are excluded. Given these representations, \sysname will produce a \textit{proximity report} (see Figure \ref{fig:reports}), summarizing the total time spent with the co-locator and the respective locations in sorted order. In a practical use-case, case investigators can consider the top N users (e.g., N = 15) with the most proximity minutes or consider specific locations suspected of at high-risk. Such strategies are already used by professional contact tracers to hone in on the most probable at-risk co-locators while eliminating users who may be false positives.

\subsubsection{Efficacy of Iterative Contact Tracing}

While the above experiment involved a single level of contact tracing, contact tracing is, in most cases, an iterative process, with each co-locator subjected to an investigation. Given that a user may come in contact with more than a hundred users in a single day (e.g., if a user attends a lecture and then visits the cafeteria), tracing for only two locations can be limiting.

We had explained how the co-locators list needs to be pruned at each step to identify the users at most risk. We had also suggested using a carefully chosen $\tau$ and $\omega$ to filter out low-risk locations. These strategies may be susceptible to missing some ``true positive'' cases. An alternative is to ``test and trace'', which combines testing with contact tracing - a strategy used by many countries for COVID-19. In this case, each co-located user is administered a test to verify their infection medically. Accordingly, only verified infected cases are subjected to iterative contact tracing while excluding others.

\begin{table*}[h]
    \centering
    \caption{Count of co-located Users by ``test and trace'' strategy}
       \begin{tabular}{lcccc} \toprule
        Contact Tracing & Round 1 & Round 2 & Round 3 & Round 4 \\ \midrule
        Selective & 275 & 438 & 764 & 1752 \\ \bottomrule
    \end{tabular}
	\label{tab:contact_tracing}
\end{table*}

With ``test and trace'', the number of users subject to contact tracing increases based on the transmission rate, R. 
For example, if R=2, then only 2 out of the several tens of users identified by our tool will be subjected to additional tracing in each step (we assume that all users are tested to find R users who are infected). Table \ref{tab:contact_tracing} depicts the number of users identified by ``test and trace''; we observe the growth is much lower than a naive iterative strategy.

\apt{With testing constrained by availability, cost, and time, the proximity report can once again be utilized to \emph{``prioritize''} and \emph{``filter''}. We prioritize by sorting the co-locators based on the amount of overlapped time duration with an infected patient and picking the top co-locators. Then, we filter to identify the users co-located with the patient for a duration of at least $\omega$ mins. Responding to COVID-19 CDC guidelines \cite{cdcDigitalTool2}, $\omega$ value is 15 minutes for the infection to spread. This procedure is performed iteratively until all high-risk co-locators are traced. Another capability is to identify at-risk co-locators by considering the number of at-risk users exposed to and exposure duration. Case investigators can carry forward these shortlisted users for manual investigation.}


\subsubsection{Contact Tracing during Quarantine Periods}
We had presented experiments during the pre-COVID semester, observing routine mobility among students and staff. Here, we examine how \sysnames's contact tracing results changes with strict lockdown policies.

\begin{figure} [h]
    {\begin{tabular}{cc}
    \includegraphics[width=2.7in,keepaspectratio]{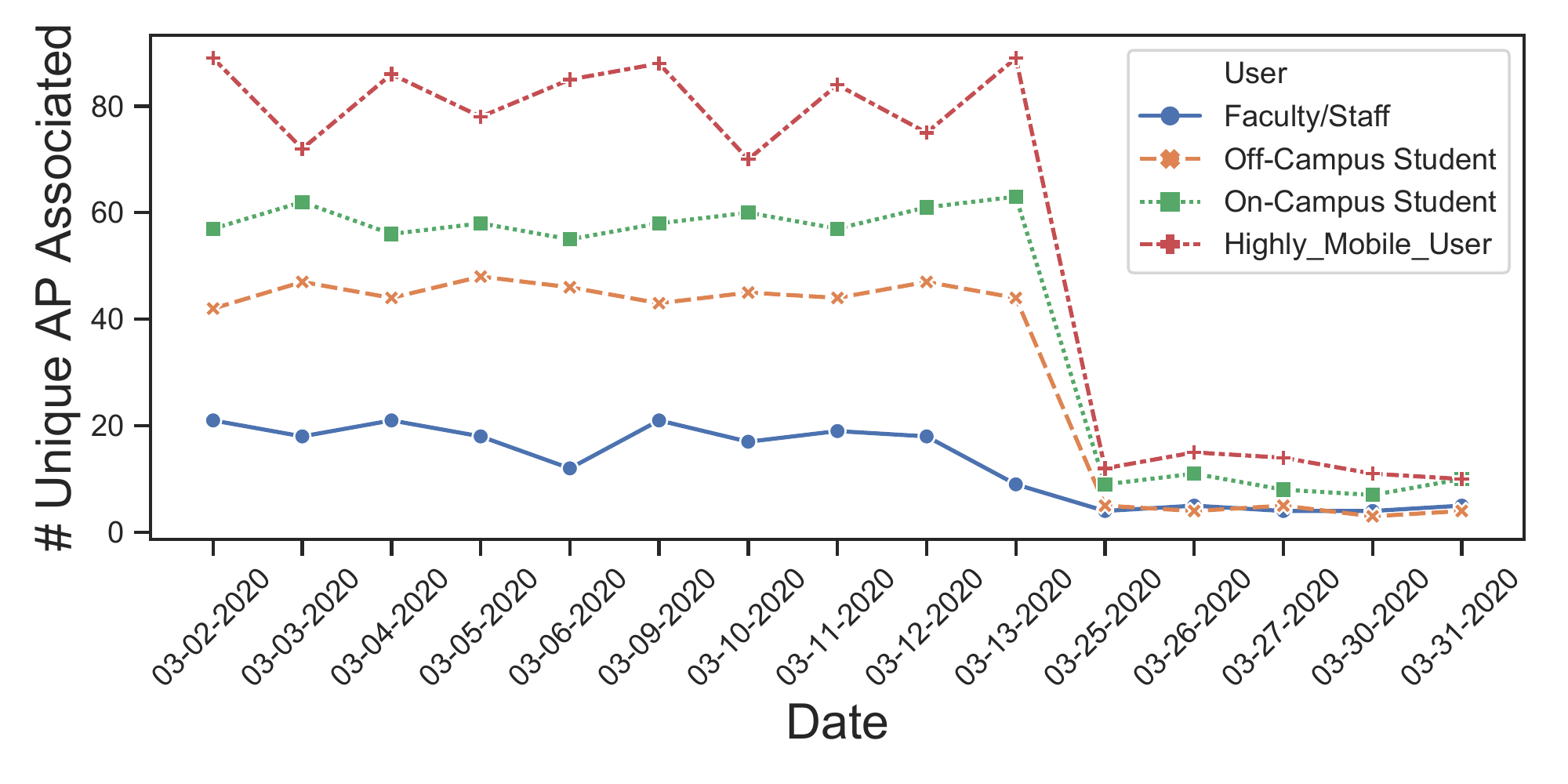} &
    \includegraphics[width=2.7in,keepaspectratio]{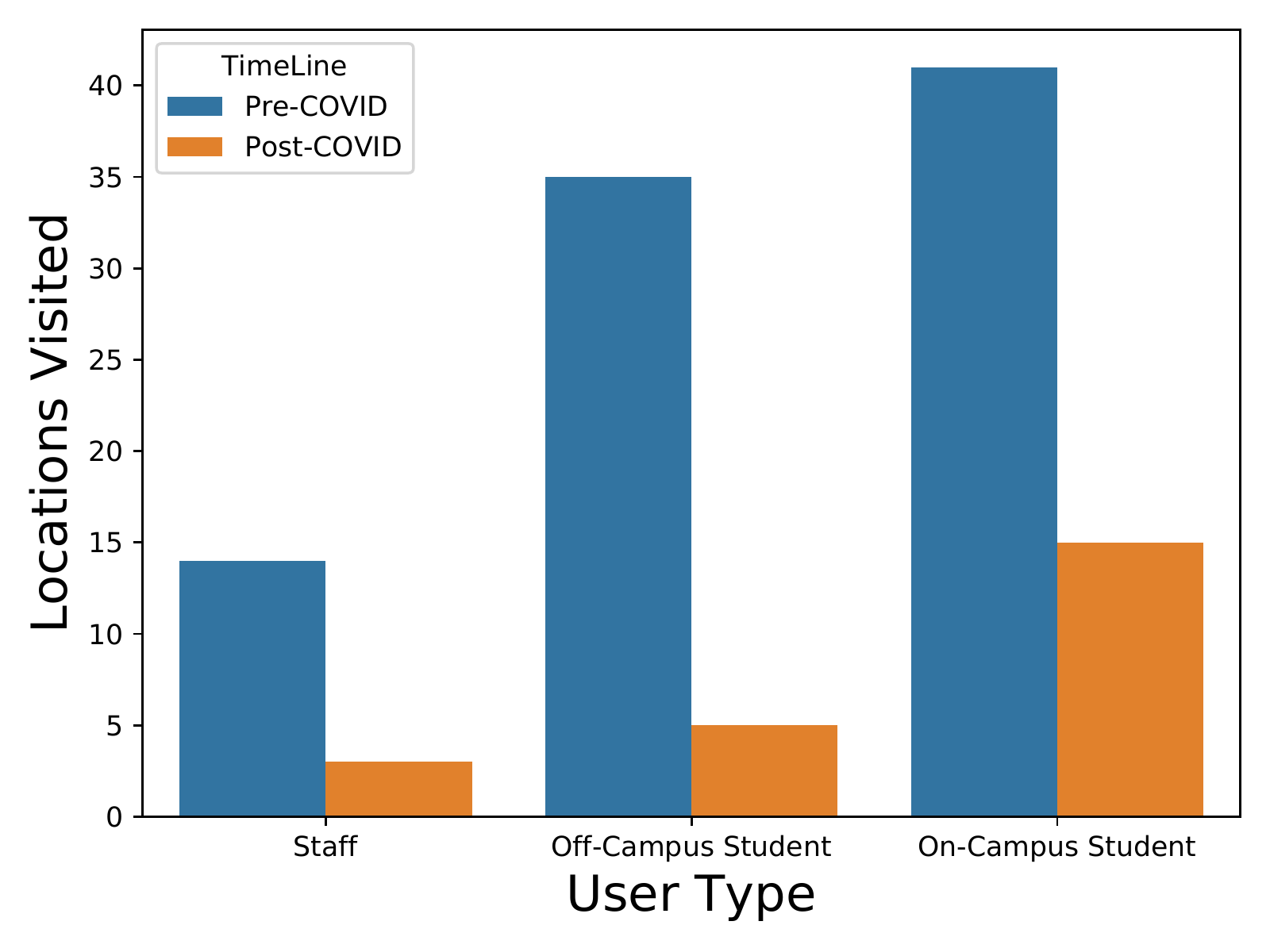}\\ 
      (a) & (b)
    \end{tabular}
    }
    \caption{(a) Number of Locations visited pre-COVID and post-COVID by 4 different user types (b) WiFi based location count for $\tau$ = 10 for different user types \final{pre COVID-19 and post COVID-19.} \label{fig:loc_count_users}}
    \vspace{-0.3cm}
\end{figure}

Figure \ref{fig:loc_count_users}(a) shows the number of locations visited by different campus user types per day. While users visited between 20-80 locations for $\tau$ = 10 mins during the normal period, we observe a sharp drop in AP location visits for all user types due to lockdown policies (after March $25^{th}$). The change in mobility will significantly alter \sysnames's contact tracing results for infected during the lockdown period. Specifically, Figure \ref{fig:loc_count_users}(b) shows the number of locations visited by one user ruled for COVID-19 contact tracing (duration of 4 days). As shown, the number of locations visited varies from 5 to 20 for $\tau$ = 10, and it drops to 1-6 locations visits for $\tau$ = 30 minutes (or greater).

\begin{table}[ht]
\begin{minipage}[c]{0.5\linewidth}
     \centering
    \includegraphics[width=0.99\columnwidth]{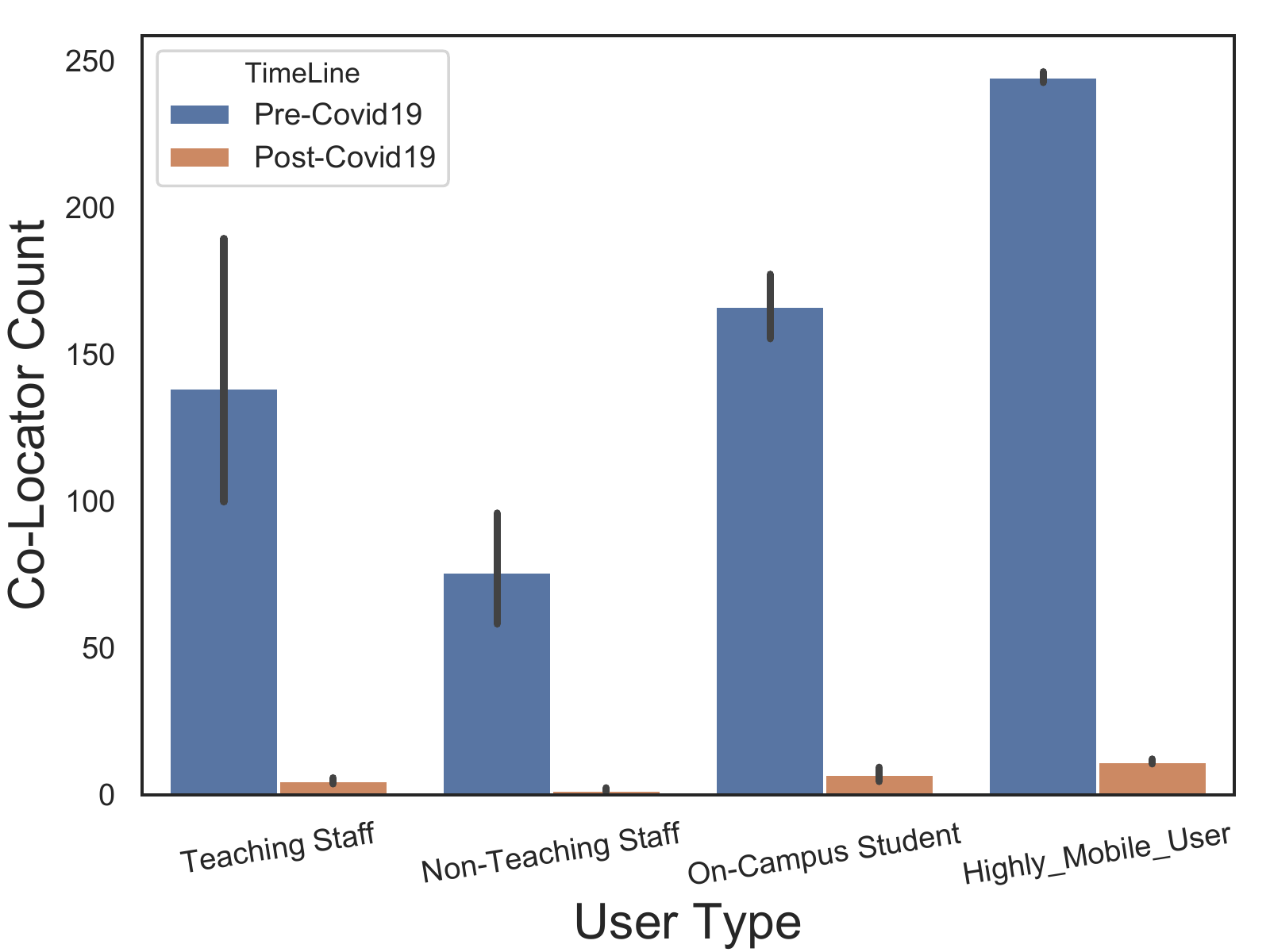}
    \captionof{figure}{Number of Co-locators  pre-COVID and post-COVID for each of the 4 different user types $\tau$ = 30mins and $\omega$ = 30mins  \label{fig:coloc_count}}
\end{minipage}
\hfill
\begin{minipage}[c]{0.45\linewidth}
     \centering
    \includegraphics[width=0.99\columnwidth]{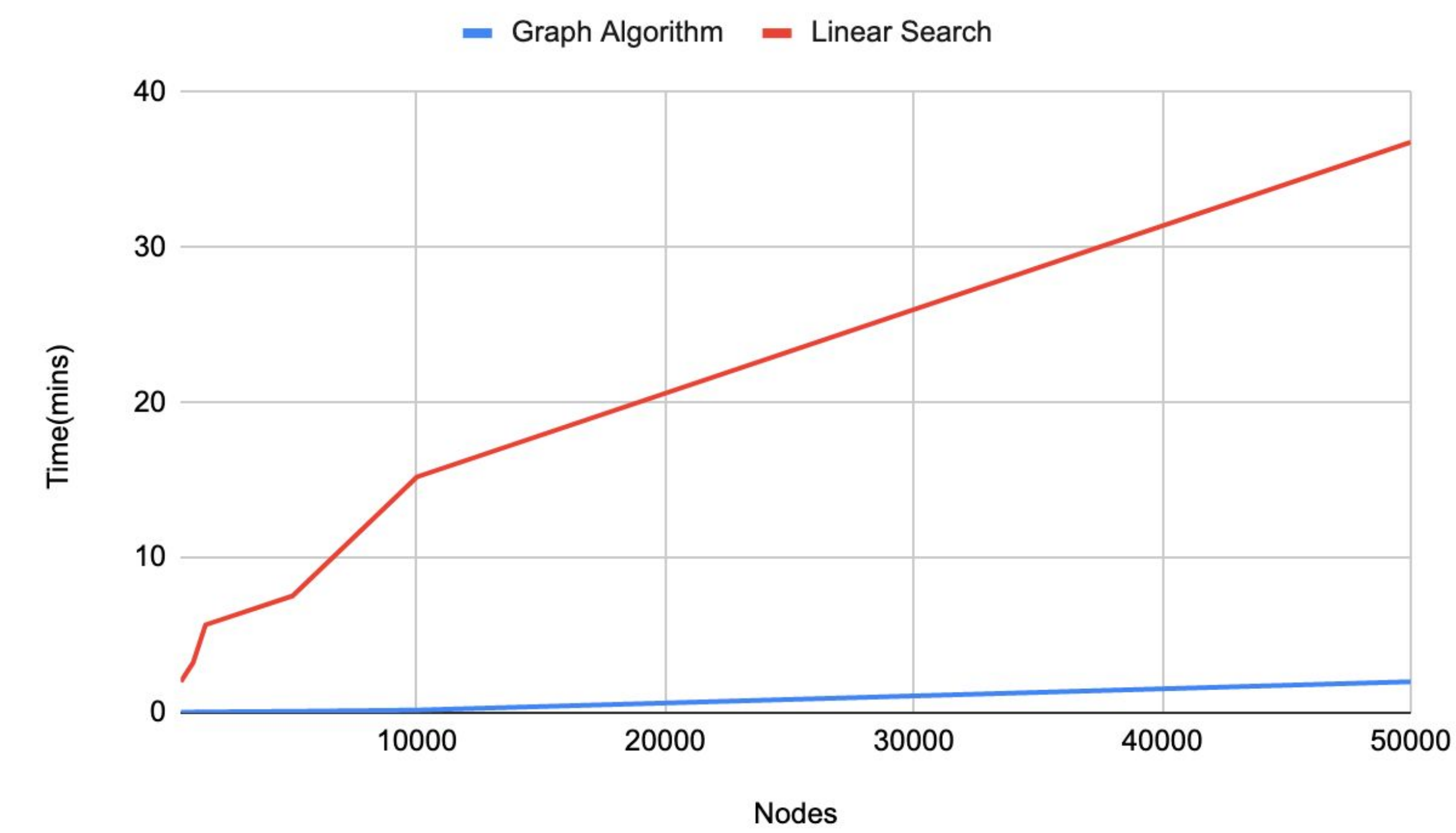}
    \captionof{figure}{Efficiency of our graph algorithm  \label{fig:graph-eval}}
\end{minipage}
\end{table}

Figure \ref{fig:coloc_count} depicts the number of co-locators for $\tau$ = 30 minutes for several users during \ncz{pre-COVID} and lockdown periods. The safety policies introduced, primarily social distancing and lockdown, have lowered the co-locator count to be less than ten for all user types, an order of magnitude reduction. In such cases, \emph{comprehensive} contract tracing of all co-locators is feasible through manual means.

\subsection{Efficacy of Our Graph Algorithm}
To evaluate the efficiency of our graph algorithm, we compare the execution time of the naive linear search approach and our graph-based algorithm across varying co-locators. Since different users display a different mobility pattern, the number of co-locators seen for each user will differ. Searching the co-locators using linear search requires a complete scan of the entire dataset sequentially, resulting in a high overhead across all runs irrespective of the number of observed co-locators of the device. Additionally, as the number of nodes increase, the search overhead increases. In contrast, our graph algorithm efficiently identifies relevant edges and nodes relevant to the specified query, thereby reducing the search space overhead. Also, adding the constraint of $\tau$ results in further pruning of edges, resulting in reduced search space, reducing our algorithm's time and space complexity.  This behavior is depicted in Figure \ref{fig:graph-eval} that compares the execution overhead of the two approaches for our campus dataset. As shown, our graph-based implementation outperforms the naive sequential search by a significant margin.

\apt{We further evaluated the graph-based algorithm with the performance of a PostgreSQL database with a single column and multiple column indexes. For the PostgreSQL database with a single index, we found the query response time for a various number of users (ranging from 5000 to 30000) across time spans of a few days to a month was similar to our solution. However, the memory overhead of the database was $\approx$ 4.5X higher 
than \sysnames.}

\apt{As we update the PostgreSQL database with daily logs to add device trajectories, the time needed to update the database indexes is in the order of \final{hours} ($\approx$1 to 5 for data from 5000 to 30000 users). Since indexes are synchronized with the tables, updating the indexes blocks further updates. We also found that indexing created an additional memory overhead of at least 250MB for a simple datastore of 20GB comprising of 5000 user trajectories for 4 weeks. This indexing overhead would only grow as the number of users and locations increases.}

\apt{We found that using multi-column indexing of the MAC and location or the MAC and date column produced higher overheads in terms of the index update time and the index memory consumption. This suggests that multi-column indexing is costlier than the single-column index. Overall, our Graph-based algorithm provides query performance similar to an indexed database without any indexing update or memory overheads.}


\subsection{Limits and Limitations of Technology}
WiFi-Sensing has well-known limitations, and this section analyzes the implications of these limitations on contact tracing.

\textbf{Multi-device Users :} Researchers have previously studied multi-device users' behavior and shown that it is prevalent among users to own two or more devices \cite{trivedi2020empirical}. A key consequence is \emph{device count seen by an AP does not equal user count}. While all WiFi logs record device association information, not all of them provide user ownership information. If such information is missing, RADIUS authentication logs should be used to map devices to owners to avoid double counting devices as separate users.

\begin{figure}[h]
\includegraphics[width=2.7in,keepaspectratio]{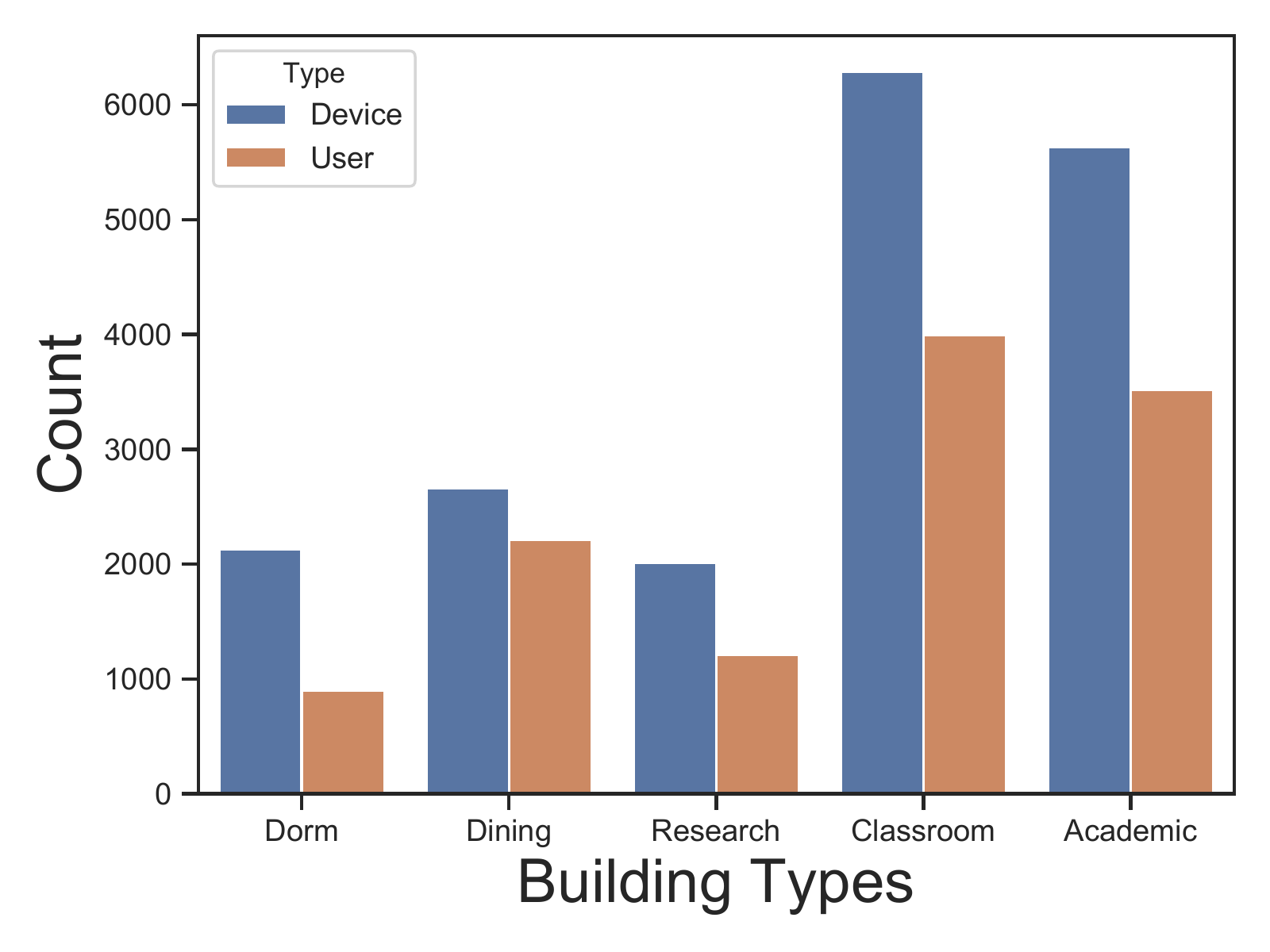}
    \caption{Unique Device and Unique User counts across buildings on a typical weekday. Need for user of device mapping information \label{fig:auth}}
\end{figure}

Figure \ref{fig:auth} shows the number of unique devices seen by APs in different campus buildings and the corresponding user count (e.g., ARUBA syslogs provide both types of information). As shown, locations such as dorms and classrooms see between 1.5X to 2X difference in unique devices and unique users (since users may connect a phone and a laptop to the network). Only dining areas (cafeteria) see low over counting since users are likely to carry only their phone when eating. This result highlights the importance of considering device ownership to avoid over counting users by only considering connected devices.

\textbf{Unassociated Devices :} Not all users may connect their mobile devices to the WiFi network. Such devices are visible to the network when they perform SSID scans using a randomized MAC address. Unassociated devices can cause multiple challenges. Ignoring them altogether will result in under-counting users. Inversely, counting all devices can yield a large number of false positives. 

\begin{figure} [h]
    {\begin{tabular}{cc}
    \includegraphics[width=2.7in,keepaspectratio]{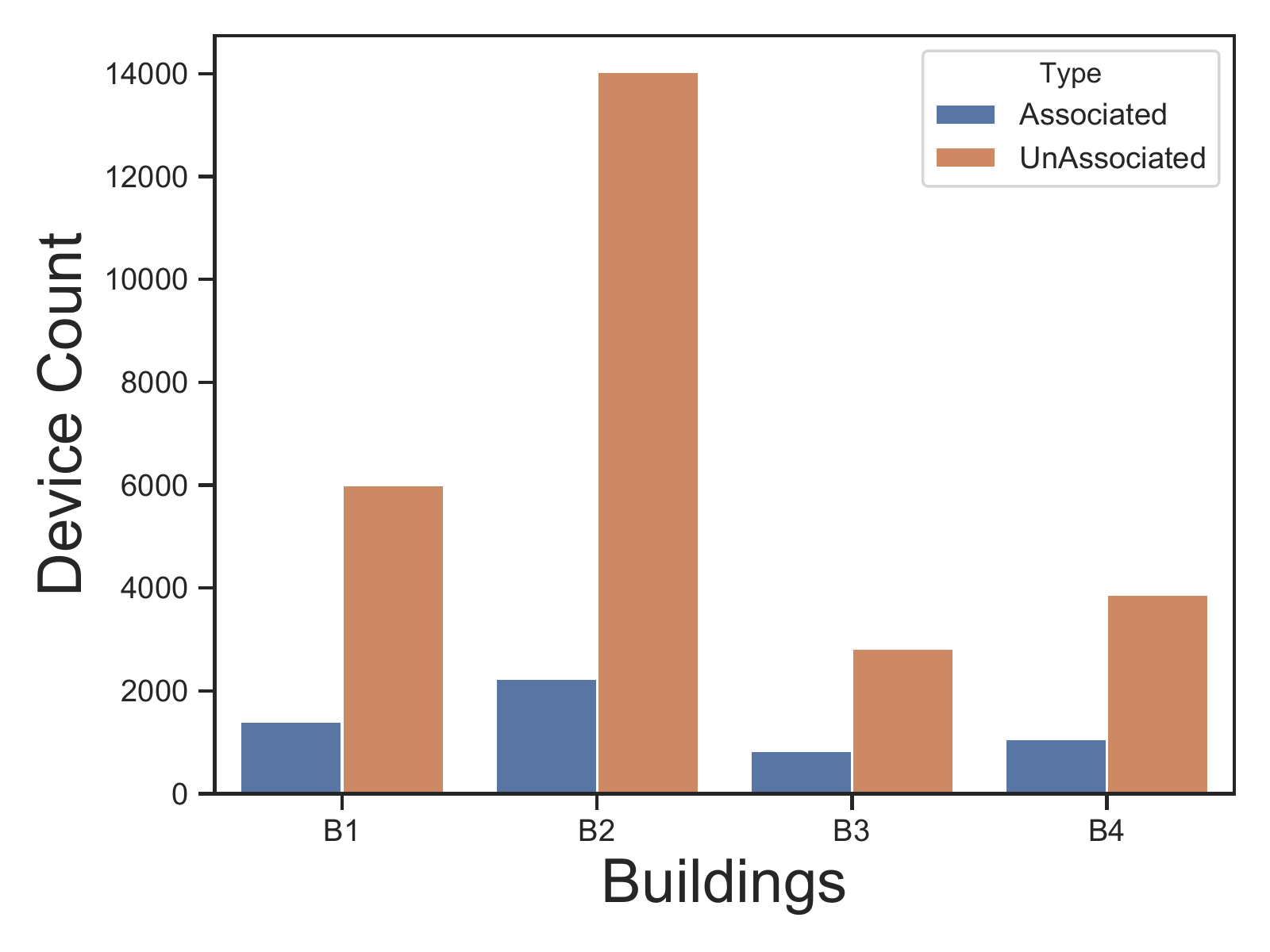} &
    \includegraphics[width=2.7in,keepaspectratio]{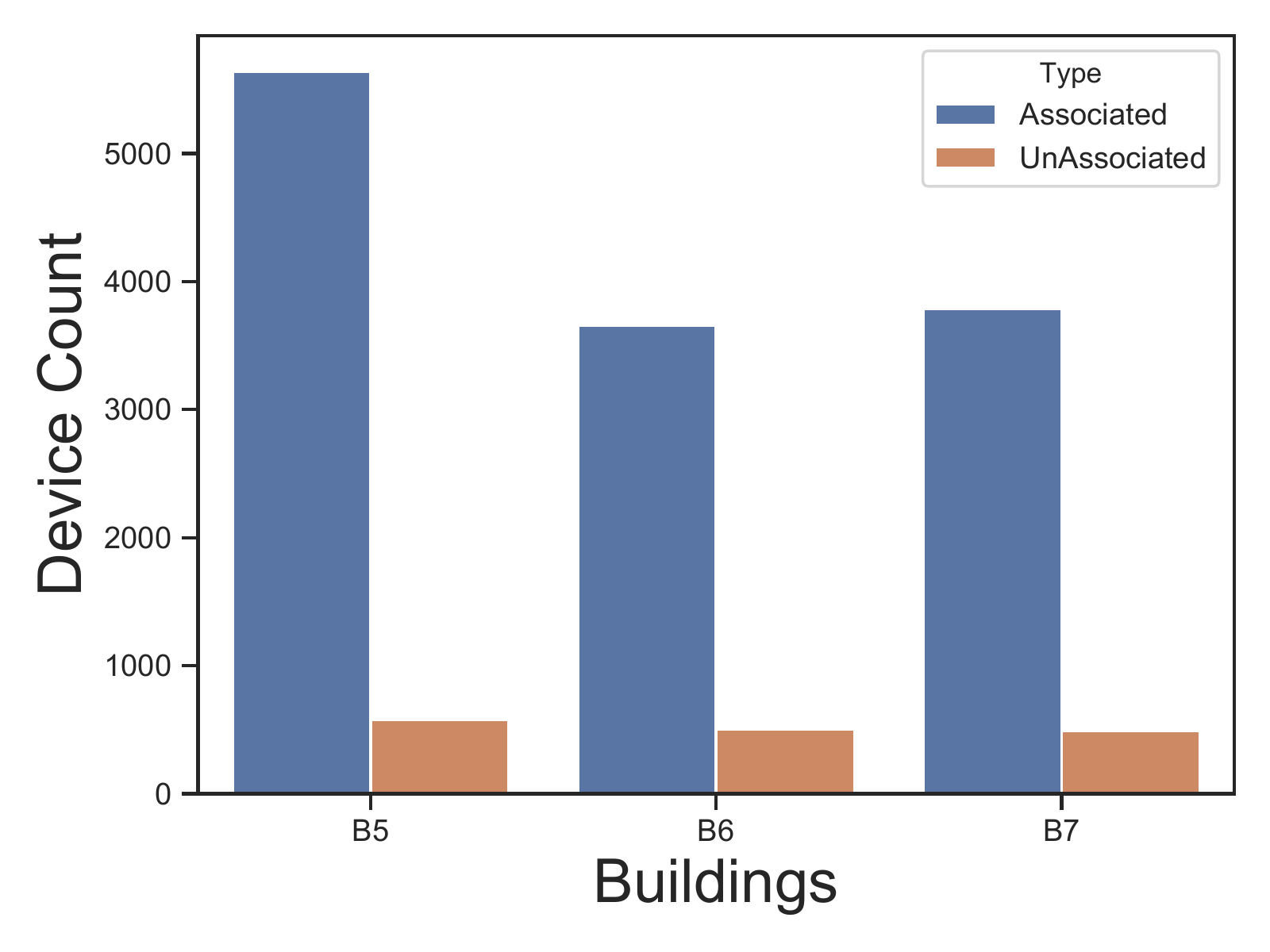}\\
    (a) & (b)
    \end{tabular}
    }
    \caption{Associated devices v/s UnAssociated devices (a) Unfiltered and (b) Filtered \label{fig:unassoc}}
    \vspace{-0.3cm}
\end{figure}

Figure \ref{fig:unassoc}(a) represents the number of unassociated devices seen in four buildings in our Singapore campus. Since the buildings are next to a public road or public bus stop, the number of unassociated devices per day is 5X greater than the number of associated users. Figure \ref{fig:unassoc}(b) shows that enforcing a session duration of 15 minutes filters out most of these \emph{chance} associations, and the number of such devices (likely visitors) is around 12\% of the total number of associated devices.

\textbf{Impact of Session Duration:} \sysname uses two parameters $\tau$ and $\omega$ that are directly related to WiFi session durations. Judicious choice of these parameters can allow for a good tradeoff between eliminating false positives and eliminating true positives.

\begin{figure} [h]
\includegraphics[width=2in,keepaspectratio]{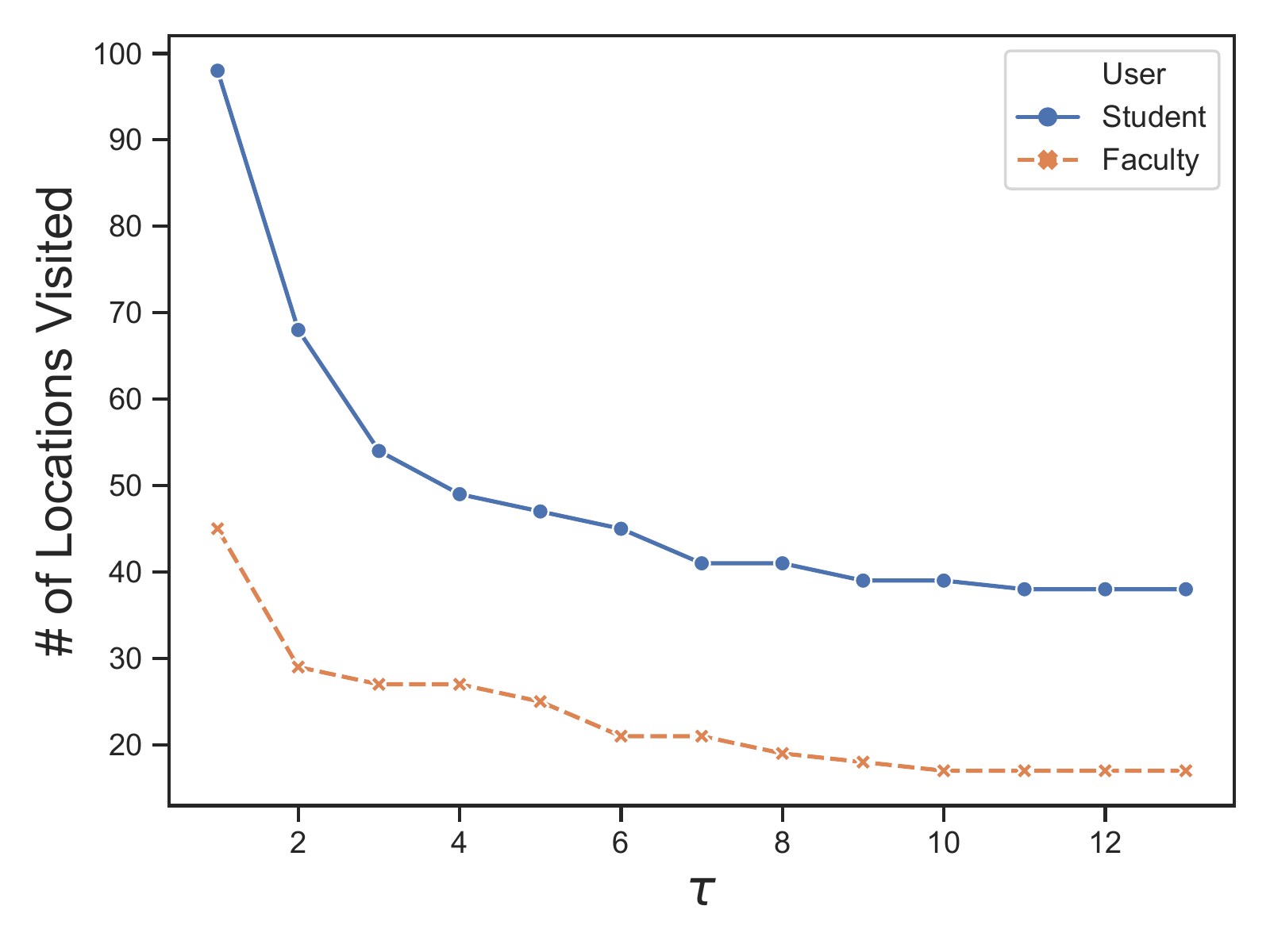}
    \caption{Count of Locations Visited by Student and Faculty\/Staff for varying values of $\tau$ \label{fig:loc_visit_T}}
\end{figure}

Figure \ref{fig:loc_visit_T} shows the number of AP locations visited by campus users for varying values of session length $\tau$. The figure shows that the location visits stabilize around $\tau$ = 10 mins and then yield 20-40 location visits per day. Small values of $\tau$ include locations visited when in transit and should be ignored.

\begin{figure} [h]
    {\begin{tabular}{ccc}
    \includegraphics[width=2.1in,keepaspectratio]{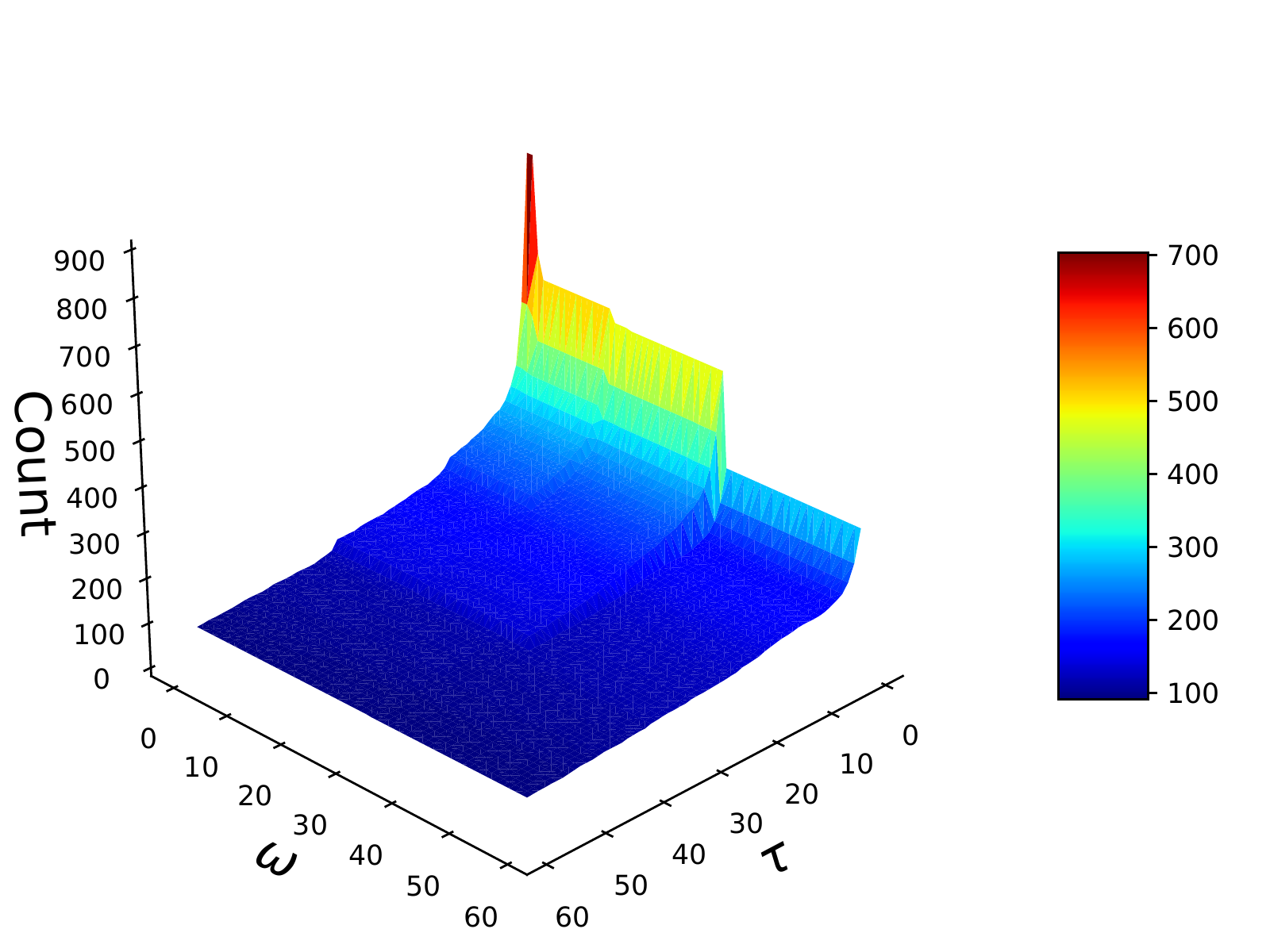} & 
    \includegraphics[width=2.1in,keepaspectratio]{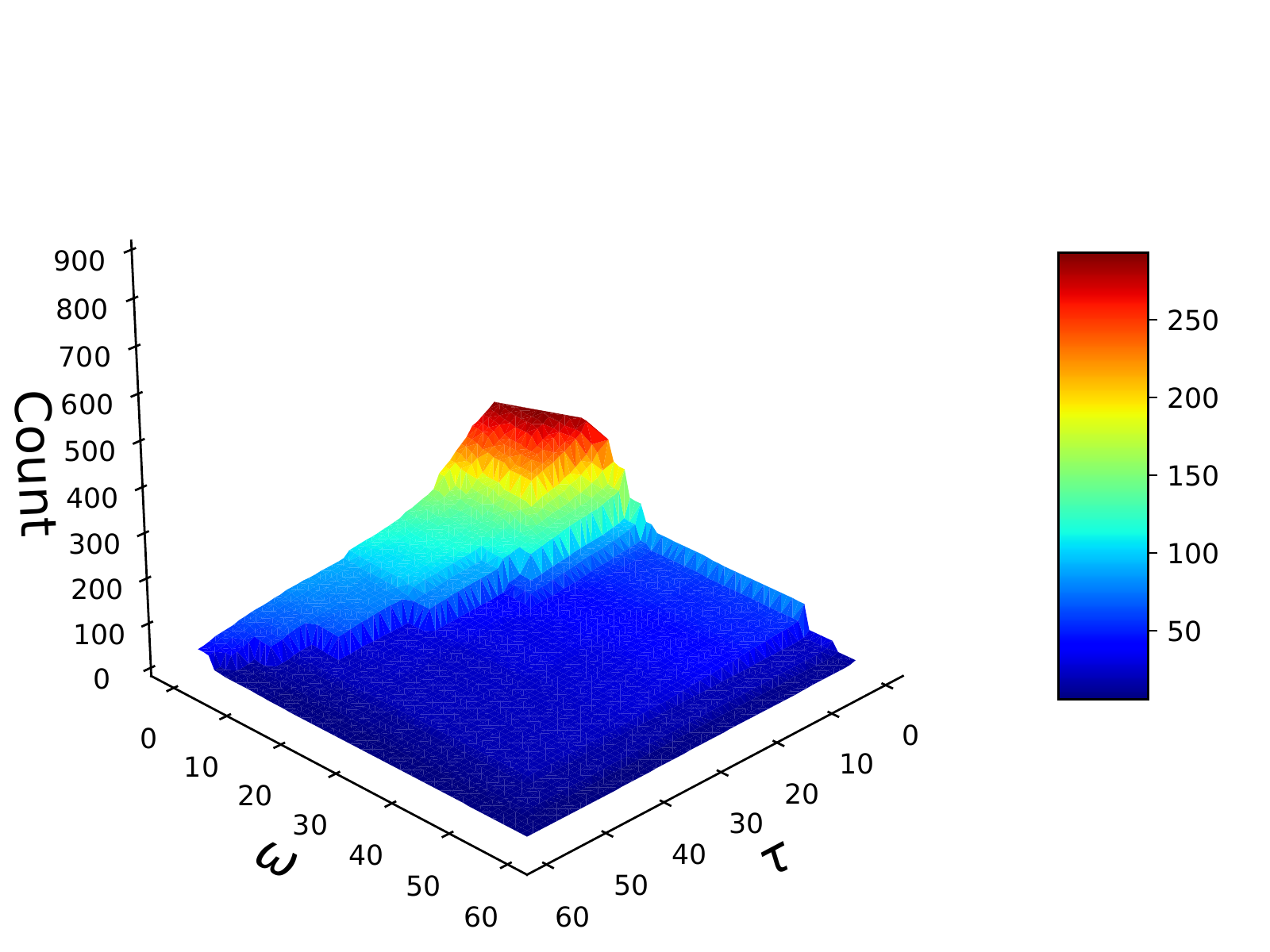} &
    \includegraphics[width=2.1in,keepaspectratio]{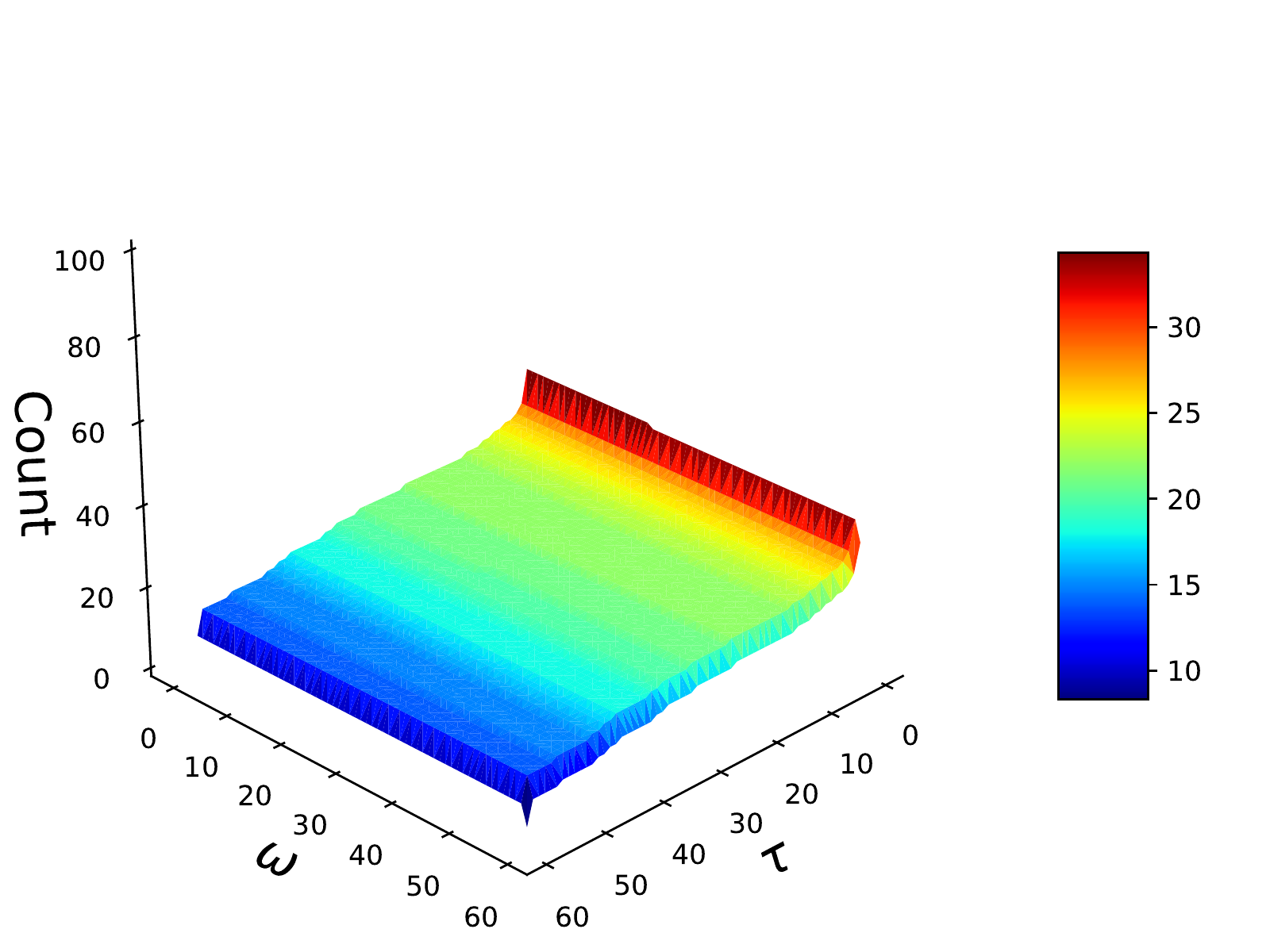} \\
    (a) & (b) & (c)
    \end{tabular}
    }
    \caption{Co-locator count for varying values of $\tau$ and  $\omega$ (a) Student (b) Teaching Faculty Staff (c) Non-Teaching Faculty Staff  \label{fig:vary}}
    \vspace{-0.3cm}
\end{figure}

\begin{figure} [h]
    {\begin{tabular}{cc}
    \includegraphics[width=1.8in,keepaspectratio]{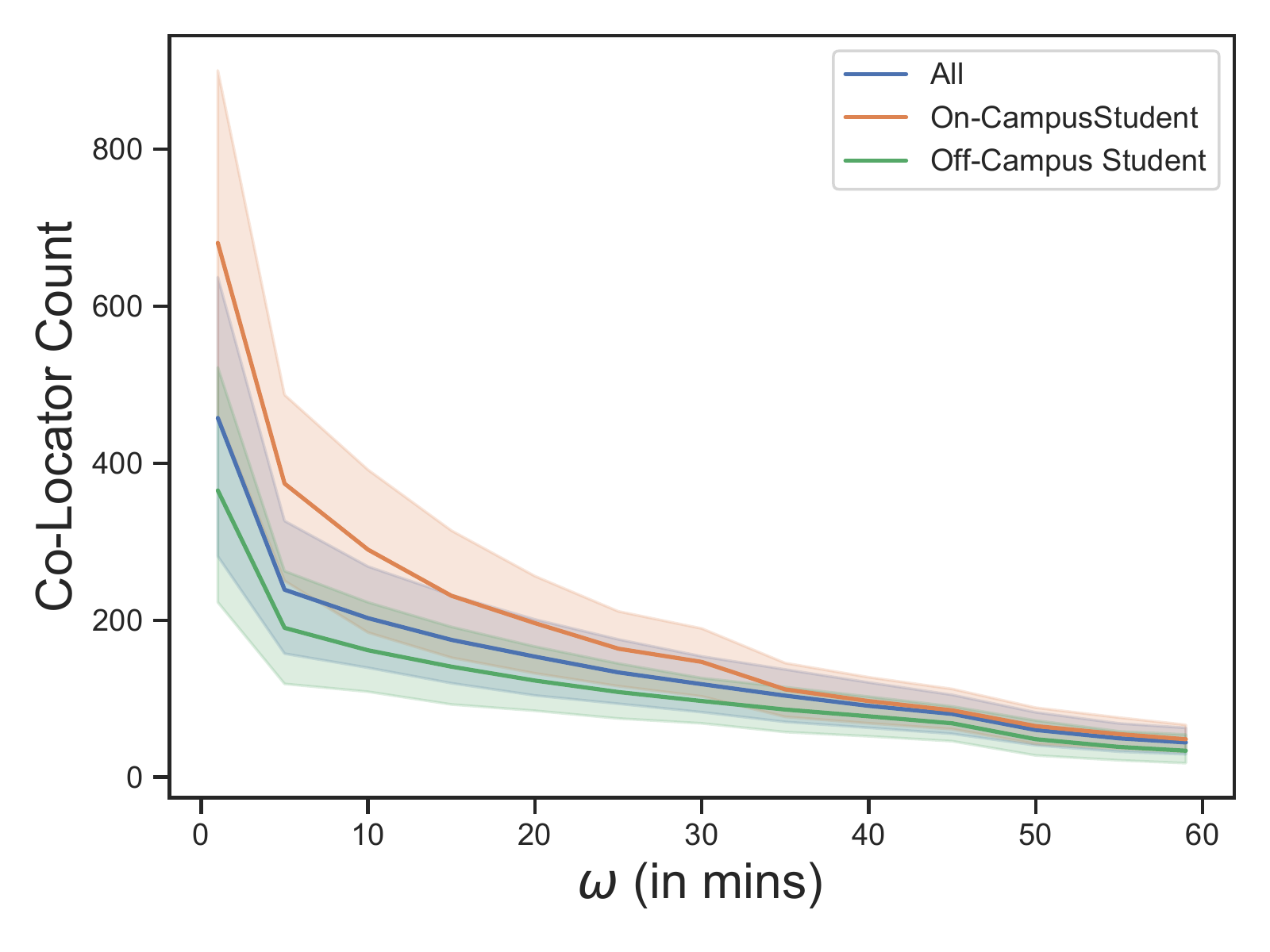} &
    \includegraphics[width=1.8in,keepaspectratio]{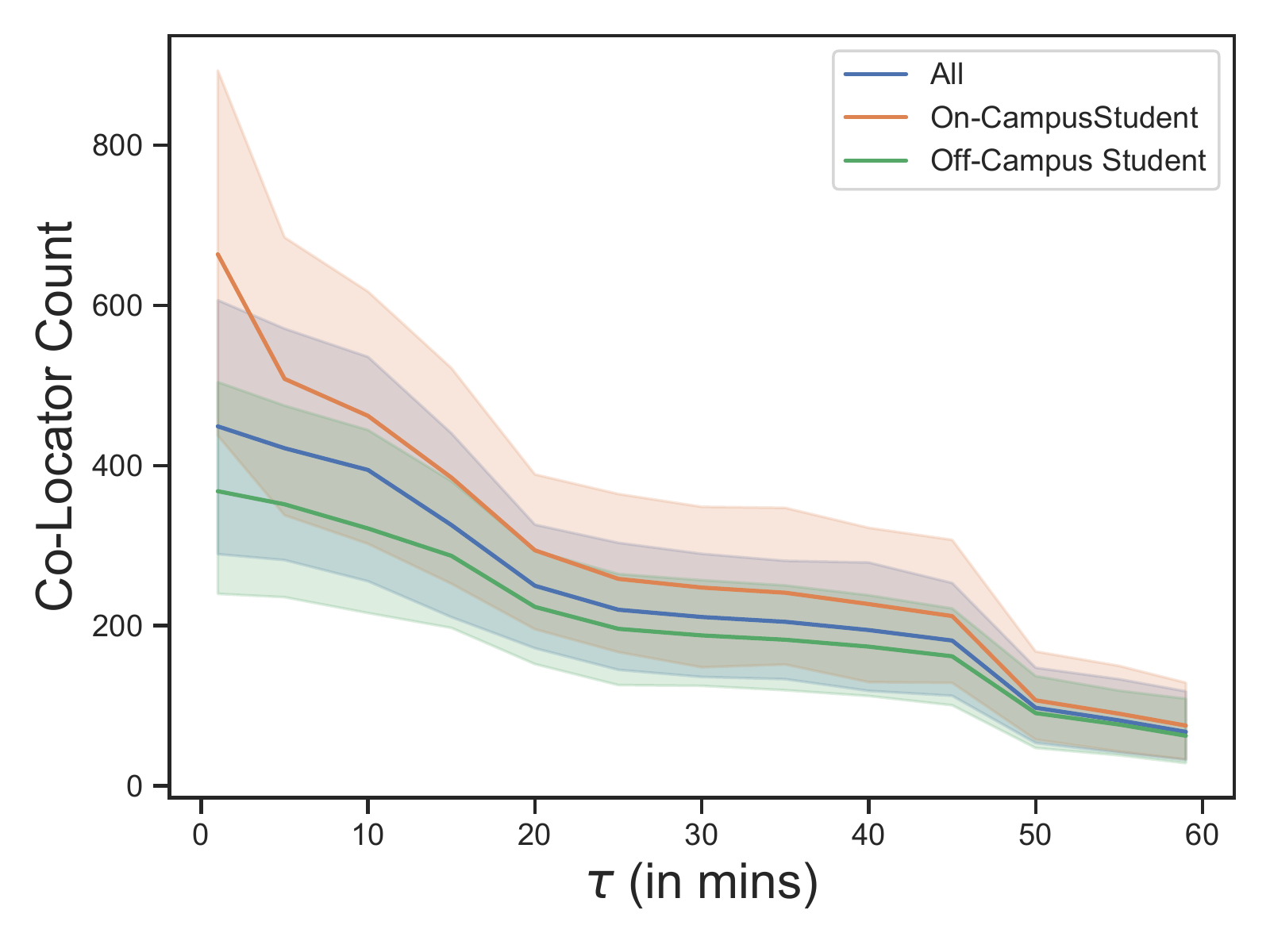}\\
    (a) & (b)
    \end{tabular}
    }
    \caption{(a) Co-locator count sliced with fixed value of $\omega$ (b) Sliced with fixed value of $\tau$ \label{fig:slice}}
    \vspace{-0.3cm}
\end{figure}

Figure \ref{fig:vary} shows the impact of varying values of $\tau$ and $\omega$, and the figure shows a decreasing gradient as both $\tau$ and $\omega$ are increased for all user types. Finally Figure \ref{fig:slice} shows the number of colocated users for varying values of $\tau$ and $\omega$. As shown, using values that are tens of minutes allows the tool to filter out overlapping sessions caused by users in transit. These results highlight the importance of carefully choosing $\tau$ and $\omega$ depending on the infectious nature of the disease and avoiding false positives.

\section{Discussions}
Here we discuss the implications and limitations of our findings.

\subsection{\apt{Supporting Case Investigations}}
\apt{At its core, contact tracing remains a labor-intensive manual task performed by qualified caseworkers to determine infected patients and disease transmission chains in the community. \sysname does not aim to replace caseworkers but instead aims to assist them by providing accurate information on possible infected individuals, locations, and contact events automatically -- without required erroneous and time-consuming interviews. \sysname was developed in collaboration with key healthcare professionals, including the director of a campus hospital and a nurse who has been running contact tracing operations for various outbreaks (such as meningitis) for many years. The tool has been carefully designed with their input to fit into their workflow. Collectively, our efforts are centered on fulfilling two key requirements for a digital contact tracing tool, as defined by the Center for Disease Control (CDC) \cite{cdcDigitalTool}:}
\begin{enumerate}
    \item \apt{\textbf{Aid the process of retrieving past location information from infected individuals:} In most cases, the use of GPS logs helps individuals accurately recall their daily activities' outdoor locations. However, GPS is unavailable indoors, and remembering all possible indoor locations visited is vital since the risk of spreading infections indoors is 20 times higher than outdoors \cite{nishiura2020closed}. \sysname greatly assists with the recall process by providing accurate proximity locations of all users, at any location (either indoors or outdoors) where WiFi is available, without requiring additional environmental instrumentation or explicit user interaction.}
    
    \item \apt{\textbf{Improve investigators' efficiency in identifying and assessing the most relevant sources of threat:} The workload on health professionals increases dramatically during pandemic times, thus reducing their contact tracing burden is essential to prevent errors due to overwork. \sysname achieves this by prioritizing highly threatening cases. It allows caseworkers to easily tweak key settings to change the thresholds used to determine the various threat levels. In particular, the thresholds specify how long a person needs to have spent in either a high-risk area or near an infected person before they are classified as ``at-risk''.}
\end{enumerate}

\subsection{\apt{Limitations and Future Work}}

\apt{WiFi network events fundamentally drive our location proximity mechanisms. This constrains our solution to areas with WiFi coverage -- most likely to be indoor spaces and a few key outdoor spaces. Enhancing \sysname to use other data sources (e.g., Bluetooth or GPS) will provide a richer source of location histories, but at the possible expense of user privacy and increased setup costs \final{(e.g.,} to install an app to collect GPS or Bluetooth scans).}


\apt{Another limitation is the inability to measure the physical distance between nearby users in proximity. \sysname can discover individuals who are likely to be at risk because they were in a compromised area. But, it cannot determine if any specific set of individuals \ncz{interacting} with each other.}

\apt{However, even this level of coarser interaction analysis is still useful as it provides an excellent first-cut of individuals who are at risk of infection. For example, in the case of COVID-19, researchers have found the virus's airborne transmission to remain viable in the air for three hours~\cite{van2020aerosol}. Hence, \sysname can detect at-risk individuals who have been in a compromised area for too long. This information can then be used by case investigators to conduct follow-up and monitoring sessions to determine the exact severity of risk for each identified individual more accurately.}

\apt{Beyond our experimental validation, more work needs to be done to identify how \sysname can support routine clinical health investigations for COVID-19 and other diseases on other campuses and in other countries. We are currently expanding our efforts to deploy \sysname to more campuses, and we welcome collaboration with any interested parties wanting to use \sysname on their campus or environment.}

\section{Related Work}
\label{sec:related}


The prevalence of many infectious diseases in our society has increased the importance of contact tracing--the process of identifying people who may have come in contact with an infected person-- for reducing its spread and disease containment \cite{taaa039,salathe2020covid}.
For performing contact tracing, the infected user needs to provide the places visited and persons in proximity or direct contact \cite{eames2003contact}. While the traditional method relies on interviews, the COVID-19 pandemic has seen using a method such as GPS, Bluetooth \cite{trace-together}, credit card records \cite{Korea}, and cellular localization.

Manual contact tracing as a mode for containment of diseases with a high transmission rate has proved to be too slow and cannot be scaled. Research \cite{nguyen2020enabling,ferretti2020quantifying,Farrahi2014} has shown that technology-aided contact tracing can aid reduce the disease transmission rate by quicker scalable tracing and help achieve quicker disease suppression. 


\subsection{BlueTooth Based Contact Tracing}

Bluetooth and Bluetooth Low Energy (BLE) based contact tracing has emerged as a possible method for proximity detection \cite{oliver2020mobile}. A handful of systems based on Bluetooth or BLE have been rolled out, few of which have been supported by the government of various countries such as Singapore \cite{trace-together} and Australia \cite{COVIDSAFE}.
The main limitation of these approaches is the need for mass adoption before it becomes effective \cite{WSJ-TT} and its reliance on Bluetooth distance measurements, which may not always be accurate.


Authenticity and privacy attacks are other key issues in using Bluetooth for contact tracing. \cite{cryptoeprint:2020:484} has shown that authenticity attacks can be easily performed on Bluetooth based contact tracing apps. Such attacks can forge the location visited and create a fake history of a user introducing risk to the society, as shown in \cite{cryptoeprint:2020:484}. Bluetooth apps suffer from privacy issues as noted in \cite{tang2020privacypreserving,raskar2020apps}. As a result, privacy issues for Bluetooth-based contact tracing has received significant attention \cite{berke2020assessing,cho2020contact, gupta2020quest}.
Privacy-preserving methods include the use of homomorphic encryption for determining contacts \cite{altuwaiyan2018epic} and private messaging to notify possible contacts \cite{xia2020return}, to name a few.
\section{Conclusions}
\label{sec:conclusions}

Technology-aided contact tracing is becoming an increasingly important tool for quick and accurate identification of co-locators. 
While the Bluetooth-based contact tracing method using phones has become popular recently, these approaches suffer from the need for a critical mass of adoption to be effective. In this paper, we presented a network-centric approach for contact tracing that relies on passive WiFi sensing with no client-side involvement. Our approach exploits WiFi network logs gathered by enterprise networks for performance and security monitoring and utilizes \ncz{them} for reconstructing device trajectories for contact tracing. Our approach is specifically designed to enhance the efficacy of traditional methods, rather than to supplant them with new technology. We presented an efficient graph algorithm to scale our approach to large networks with tens of thousands of users. We implemented a full prototype of our system and deployed it on two large university campuses. We \final{validated} our approach and demonstrate its efficacy using case studies and detailed experiments using real-world WiFi datasets. Finally, we discussed the limitations and privacy concerns of our work and have made our source code available to other researchers under an open-source license ~\footnote{We have released WiFiTrace under BSD for use by other parties at \url{http://wifitrace.github.io/}}.

\section*{Acknowledgements:} We thank the anonymous reviewers for their insightful comments that improved the paper. This research is supported by NSF grants 1763834 and 1836752.

\bibliographystyle{ACM-Reference-Format}
\bibliography{main}

\end{document}